\documentclass[final,3p,times,twocolumn]{elsarticle}
\usepackage{graphicx,times}
\usepackage{amssymb,amsmath}
\usepackage{aas_macros}
\usepackage{pdflscape}

\usepackage{threeparttable}
\usepackage{supertabular}

\usepackage{xcolor}

\usepackage{amssymb}
\usepackage{adjustbox}
\usepackage{subfigure}
\usepackage{multirow}
\usepackage{threeparttable}
\journal{Journal of High Energy Astrophysics}

\begin{document}

\begin{frontmatter}

\title{Measurements of spin and orbital parameters in Cen X-3 by Insight-HXMT}

\author[label1,label2]{Qi~Liu}
\author[label1,label2]{Wei~Wang}
\ead{wangwei2017@whu.edu.cn}

\address[label1]{Department of Astronomy, School of Physics and Technology, Wuhan University, Wuhan 430072, China}
\address[label2]{WHU-NAOC Joint Center for Astronomy, Wuhan University, Wuhan 430072, China}


\begin{abstract}
We present a detailed temporal analysis for the eclipsing high-mass X-ray binary system Cen X-3 using the Insight-HXMT data in 2018 and 2020. Three consecutive and high statistic observations among data are used for the precise timing analysis. The pulse profiles are revealed to vary with energy and time. The pulse profiles for the 2018 observations showed a double peak in the low energy bands below 10 keV and evolved to a single peak in higher energies without the correlation between pulse fraction and flux, and profiles in low energies changed with time. But the pulse profile for the 2020 observation only showed a broad single-peaked pulse in all energy bands with a positive relation between pulse fraction and flux, which may indicate the transition of the emission patterns from a mixture of a pencil and a fan beam to a dominated pencil-like beam. With performing a binary orbital fitting of spin periods, we obtain an accurate value for the spin period and the orbital parameters. The intrinsic spin period of the neutron star is found to be $4.79920 \pm 0.00006$ s at MJD 58852.697, with the orbital period determined at $P_{\rm orb}=2.08695634\pm 0.00000001$ day, and its decay rate of -(1.7832 $\pm$ 0.0001) $\times$ 10$^{-6}$ yr$^{-1}$ for the binary. 
\end{abstract}

\begin{keyword}
stars: neutron - pulsars: individual: Cen X-3 - X-rays: binaries 
\end{keyword}

\end{frontmatter}


\section{Introduction}
\label{sec:introduction}

High-mass X-ray binaries (HMXBs) comprised of a neutron star (NS) and a massive companion star, are one of the brightest X-ray sources in our Galaxy. Due to the strong gravity, the falling gas from the companion is accreted into the neutron star. Through the magnetosphere, the accreted matter is funneled to the magnetic polar caps of the NS magnetic field lines and forms the accretion columns above it. The X-ray emissions can be produced through the shocks in the accretion columns and escape to form the observed pulsed emissions. HMXBs can be divided into Be/X-ray binary and supergiant X-ray binary (SgXB). In the Be/X-ray binary systems, the X-ray emission is generated by the accretion of the circumstellar material. In the supergiant X-ray binaries, the neutron star can either accrete matter through the fast stellar wind or Roche lobe overflow. In Roche-lobe accreting systems, the falling materials from the companion carry the angular momentum, and generally form the accretion disc around the NS. The disk accretion systems typically possess a relatively high luminosity ($\sim 10^{38}$ erg\ s$^{-1}$). 

The neutron star X-ray binary Cen X-3, first discovered by Uhuru \citep{1971ApJ...167L..67G,1972ApJ...172L..79S}, is an eclipsing high-mass X-ray binary. Cen X-3 harbors a neutron star with a spin period of $\sim$4.8 s with an orbital period of $\sim$2.1 days with an orbital decay rate of $\sim$1.799$ \ \times 10^{-6}$ yr$^{-1}$ \citep{2015AA...577A.130F}. The neutron star has been measured with a mass of $\sim$1.21 $\pm$ 0.21 $M_{\odot}$ at the estimated distance of $\sim$ 6.9 kpc based on Gaia measurements \citep{2021MNRAS.507.3899V}. And the optical companion in the pulsar binary is an O6-8 III supergiant star \citep{1974ApJ...192L.135K,1979ApJ...229.1079H}, with a measured mass of $\sim$20.5 $\pm$ 0.7 $M_{\odot}$ and radius of 12 $R_{\odot}$ \citep{1999MNRAS.307..357A}.  Cen X-3 exhibits variations in their pulse profiles, the cyclotron resonance scattering features (CRSF, \citealt{1998A&A...340L..55S,2000ApJ...530..429B,2008ApJ...675.1487S} , and the quasi-periodic oscillations observed at $\sim 40$ mHz \citep{2008ApJ...685.1109R,1991PASJ...43L..43T,2022MNRAS.516.5579L}. The relative bright X-ray luminosity ($\sim5 \times 10^{37}$ erg\ s$^{-1}$, \citealt{2008ApJ...675.1487S}) and a secular spin-up trend in this system may suggest the existence of the accretion disc. 

Among the HMXBs, only a few give rise to X-ray eclipses (e.g., Vela X-1, \citealt{2014MNRAS.440.1114W}; Cen X-3 \citealt{2015AA...577A.130F}). The X-ray eclipses in Cen X-3 last about 20\% of the orbit. Therefore, these systems provide a possibility of accurate measurements for the orbital parameters, such as the orbital period. Since Cen X-3 is also a persistent source, it is a very good candidate for the studies of accurately measuring their orbital parameters. Cen X-3 is known to show a secular decay in the orbital period, which might be ascribed to tidal effect \citep{2000ApJ...530..429B}. The orbital ephemeris of Cen X-3 has been studied (see \citealt{1992ApJ...396..147N,2010MNRAS.401.1532R,2015AA...577A.130F}). We would expect to further reveal the orbital period as well as its changes with recent years' observations performed by Insight-HXMT. 

The Insight-HXMT launched on 15th June 2017, is China's first X-ray astronomical satellite. It can obtain good-quality timing and light curves for X-ray sources with its broad energy band of 1-250 keV for studying compact objects. The observations of Cen X-3 can provide accurate measurements of orbital parameters for the binary system. In this paper, we report the temporal analysis results for the persistent source Cen X-3 based on the available long-term monitoring observations carried out with Insight-HXMT. 

The paper is organized as follows. In Section 2, we briefly describe the HXMT observations and data reduction. The energy-dependent pulse profiles and the evolution of the profiles during the observations are presented in Section 3. The results for the measurements of the spin and orbital parameters along with the orbital period decay are discussed in Section 4. The conclusion is summarized in Section 5.

\section{Observations and Data reduction} \label{sec:Data}

To accurately measure the spin period and orbital parameters, we use three observations of Cen X-3 performed by Insight-HXMT (\citealt{2020SCPMA..6349502Z}) in 2018 and 2020 due to long exposure intervals, with a total exposure of $\sim$90 ks. The detailed information is listed in Table \ \ref{tab:ObsIDs}. Insight-HXMT satellite contains three scientific payloads, the High Energy X-ray telescope (HE; 20-250 keV), the Medium Energy X-ray telescope (ME; 5–30 keV), and the Low Energy X-ray telescope (LE; 1–15 keV) with the effective areas of 5000 $\rm cm^2$, 952 $\rm cm^2$, and 384 $\rm cm^2$, respectively. The time resolution of Insight-HXMT is 25 $\mathrm{\mu} s$, 276 $\rm \mu s$, and 1 ms for HE, ME, and LE, respectively.

To get the scientific products, we first perform the data reduction, and there Insight-HXMT Data Analysis Software (HXMTDAS v2.04) is adopted. The analysis methods are also demonstrated in previous publications (see \citealt{2021JHEAp..30....1W,2022MNRAS.513.4875C}). Here we briefly summarize the main procedures with the following standard criteria. We use tasks or commands \textsl{he/me/lepical} to calibrate from the events files. Tasks \textsl{he/me/legtigen} are used to select a good time interval (GTI) when the pointing offset angle is less than $0.04^\circ$, the elevation angle is greater than $10^\circ$, the geomagnetic cut-off rigidity is greater than 8 GeV, and the time intervals is without the South Atlantic Anomaly passage. Tasks \textsl{he/me/lescreen} are used to screen the good events based on the GTI. We used commands \textsl{he/me/lelcgen} to extract the light curves with a time resolution of 0.0078125 (1/128) sec in the energy bands 30–100 keV, 10–30 keV, and 1–10 keV for HE, ME, and LE, respectively.

\begin{table}
    \centering
    \footnotesize
    \caption{The observation IDs of Insight-HXMT used in this study.}
    \label{tab:ObsIDs}
    \begin{tabular}{l|cccc}
    \hline \hline 
    \multirow{2}{*}{ObsID} & \multirow{2}{*}{Time Start [UTC]}  & \multirow{2}{*}{Exposure time [s]} & \multirow{2}{*}{MJD}\\ \\
    \hline 
    P0101311008 & 2018-02-04 & 54239 & 58153\\
    P0201012324 & 2020-01-05 & 19110 & 58853\\
    P0201012325 & 2020-01-17 & 17010 & 58865 \\
    \hline
    \end{tabular} 
\end{table}

\begin{table}
    \centering
    \footnotesize
    \caption{The duration of each observational segment (ExpID) for three observations. A few of ExpIDs are missing in the table because the data is not available or the spin period is not detected.}
    \label{tab:ExpIDs}
    \begin{tabular}{l|cc}
    \hline \hline 
    \multirow{2}{*}{ObsID} & \multirow{2}{*}{ExpID}  & \multirow{2}{*}{Duration (s)} \\ \\
    \hline 
    P0101311008 & 01 & 11790. \\
    & 04 &  9510.  \\
    & 05 &  9060.  \\
    & 06 &  31320. \\
    & 07 &  7170.  \\
    & 08 &  9030.  \\
    & 09 &  9990.  \\
    & 10 &  14910. \\ \hline
    P0201012324 & 01 & 9600. \\
    & 02 &  9540.  \\
    & 06 &  8970.  \\
    & 07 &  6780.  \\
    & 08 &  6780.  \\
    & 09 &  8970. \\
    & 10 &  8970.  \\
    & 11 &  8880.  \\ \hline
    P0201012325 & 01 & 12810.  \\
    & 02 &  7200.  \\
    & 03 &  7230.  \\
    & 04 &  9120.  \\
    & 05 &  9030.  \\
    & 06 &  8670.  \\
    & 07 &  11520.  \\
    \hline
    \end{tabular} 
\end{table}

\begin{figure}
    \centering
    \includegraphics[width=.5\textwidth]{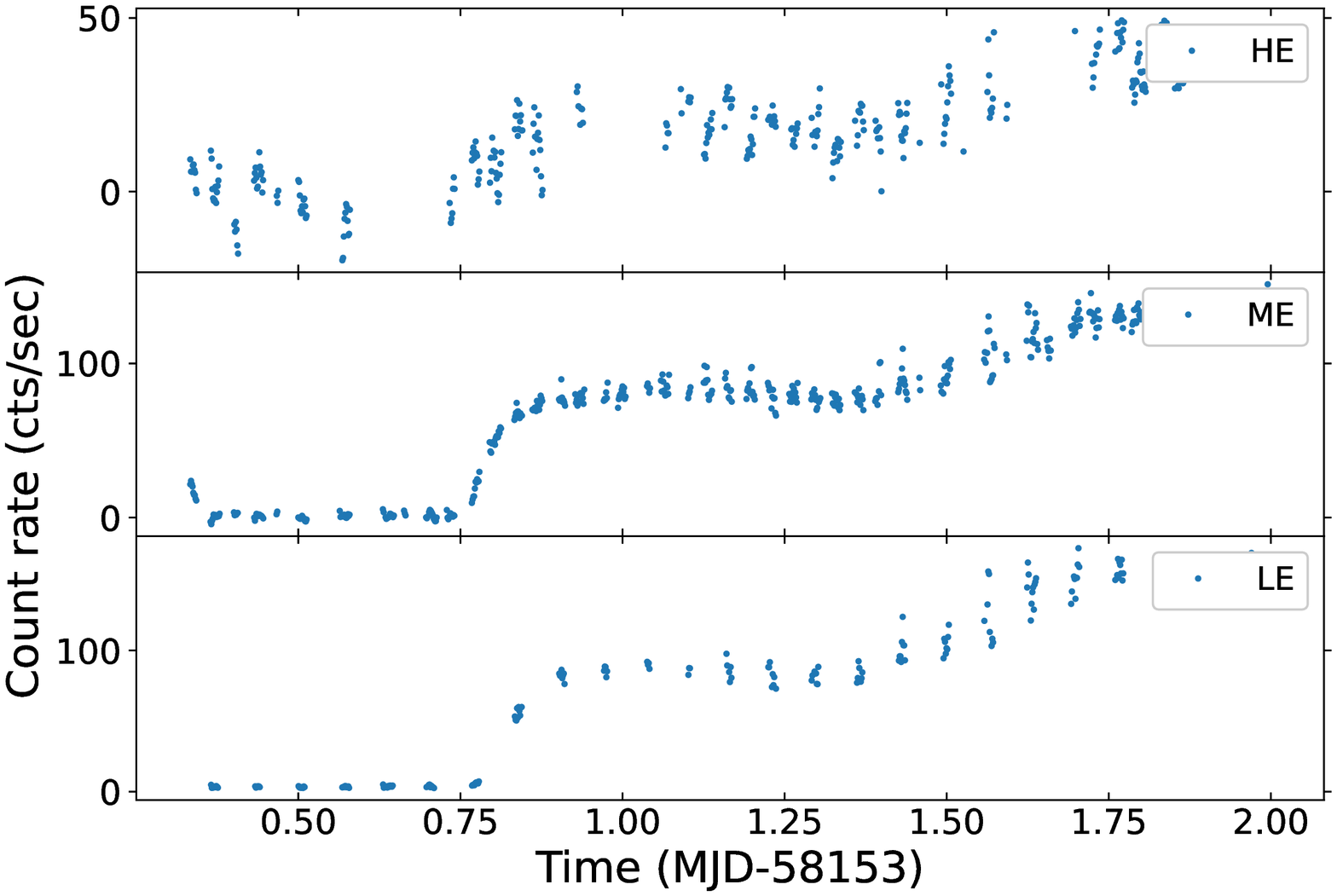}
    \includegraphics[width=.5\textwidth]{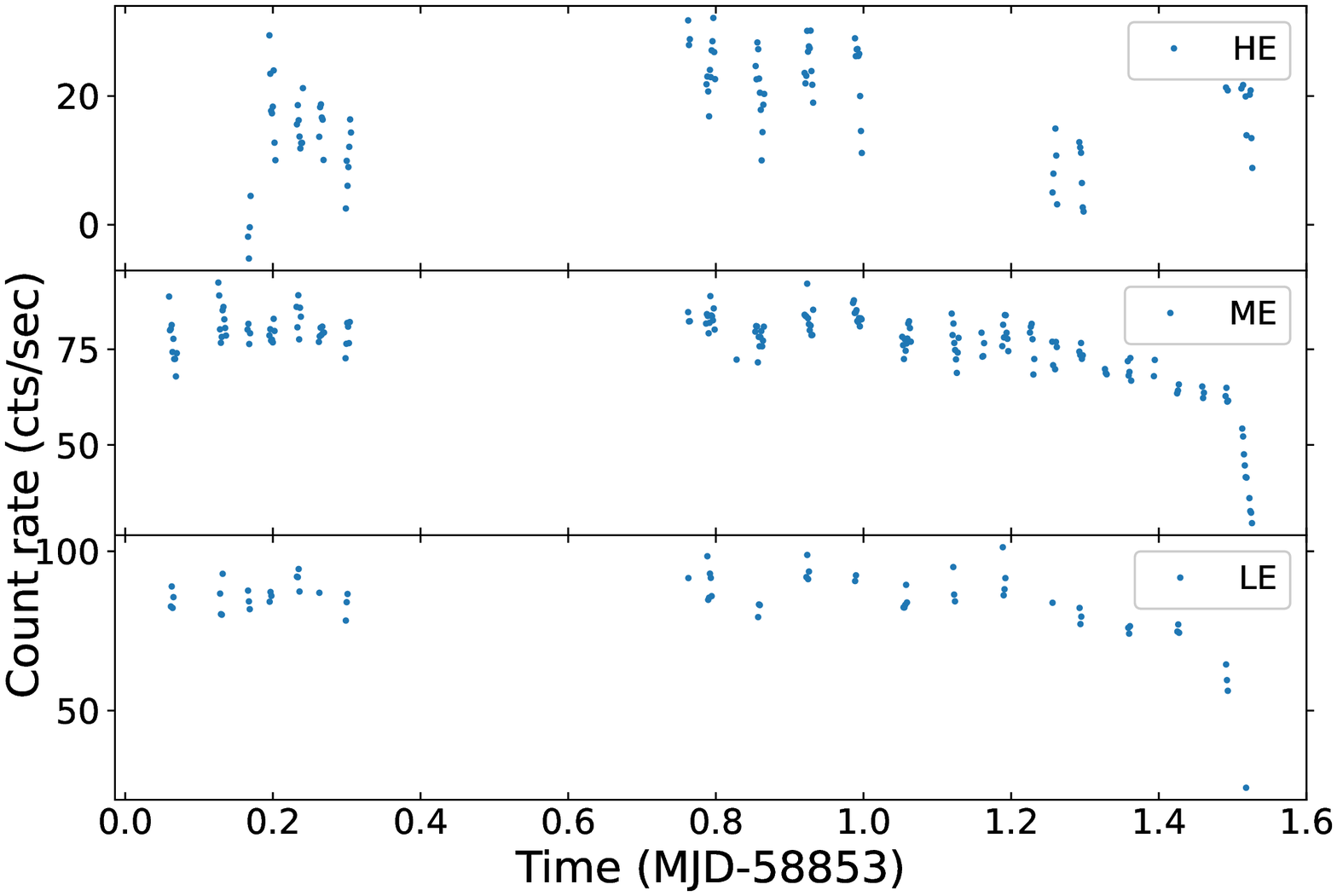}
    \includegraphics[width=.5\textwidth]{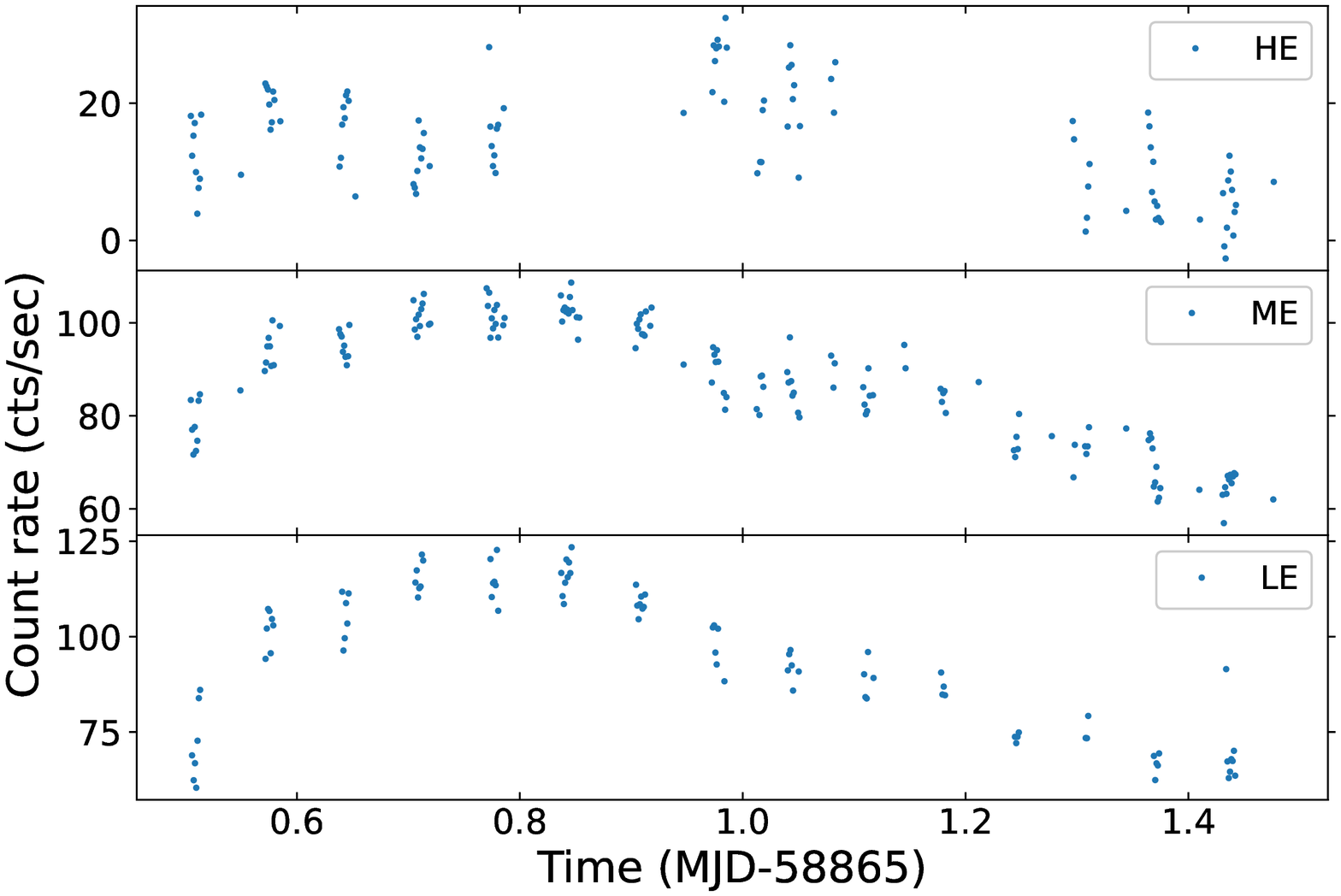}
    \caption{The figures show X-ray light curves with the background subtracted and a time resolution of 100 s for three observations in 2018 and 2020 for LE (1-10 keV), ME (10-30 keV), and HE (30-100 keV), respectively.}
    \label{fig:lc}
\end{figure}

\section{Timing Analysis} \label{sec:result}

\subsection{The pulse profiles}

The previous studies have shown the count rate variations for all available observations from 2017 to 2020 (see the upper panel of Figure 1 in \citealt{2022MNRAS.516.5579L}). In this work, we will select three observations with multiple time intervals in 2018 and 2020 to probe the orbital variation of the system. Figure\ \ref{fig:lc} shows the light curves of Cen X-3 during three observations with a time resolution of 100 s. Each observation was divided into several consecutive segments (also called 'exposure ID') recommended by the Insight-HXMT group (see Table \ \ref{tab:ExpIDs}). To perform a temporal analysis, the arrival times of events were corrected to the solar system barycenter using task \textsl{hxbary}. Using the epoch folding method in \textsl{stingray} python package \citep{2019ApJ...881...39H}, we search for the pulsations by folding the light curve in the energy range of 10-30 keV with 50 phase bins/period over a series of periods around an initial estimation for the spin period of $\sim$ 4.799 s. The frequency step for Cen X-3 is set to be less than about 10 times of the frequency resolution df=1/length Hz. We can evaluate the pulse period by the standard chi-squared statistics $\chi^2_{\rm nbin-1}$ and take the peak as the best value under a detection level of 99.9\% \citep{1983ApJ...272..256L}. Here the 1 $\sigma$ uncertainty of the measured pulse period can be estimated from the change of $\chi^2$ distribution \citep{2001ApJ...554L.177Z}. 

\begin{figure}
    \centering
    \includegraphics[width=.5\textwidth]{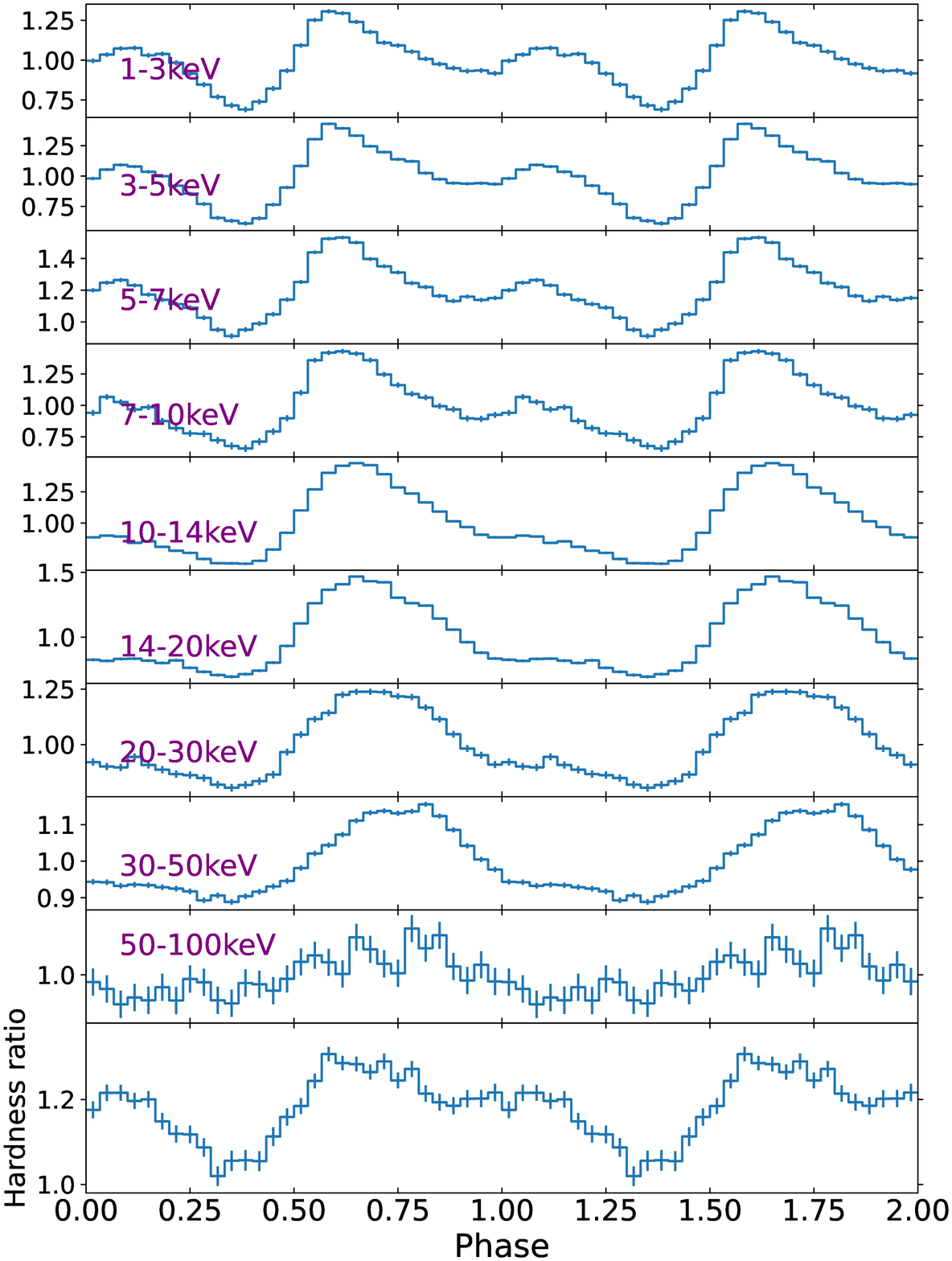}
    \caption{The pulse profiles of Cen X-3 for the ExpID P010131100809 (MJD-58154) in different energy bands. The bottom panel shows the hardness ratio of 3-5 keV/1-3 keV.}
    \label{fig:pulse}
\end{figure}

\begin{figure*}
    \centering
    \includegraphics[width=.3\textwidth]{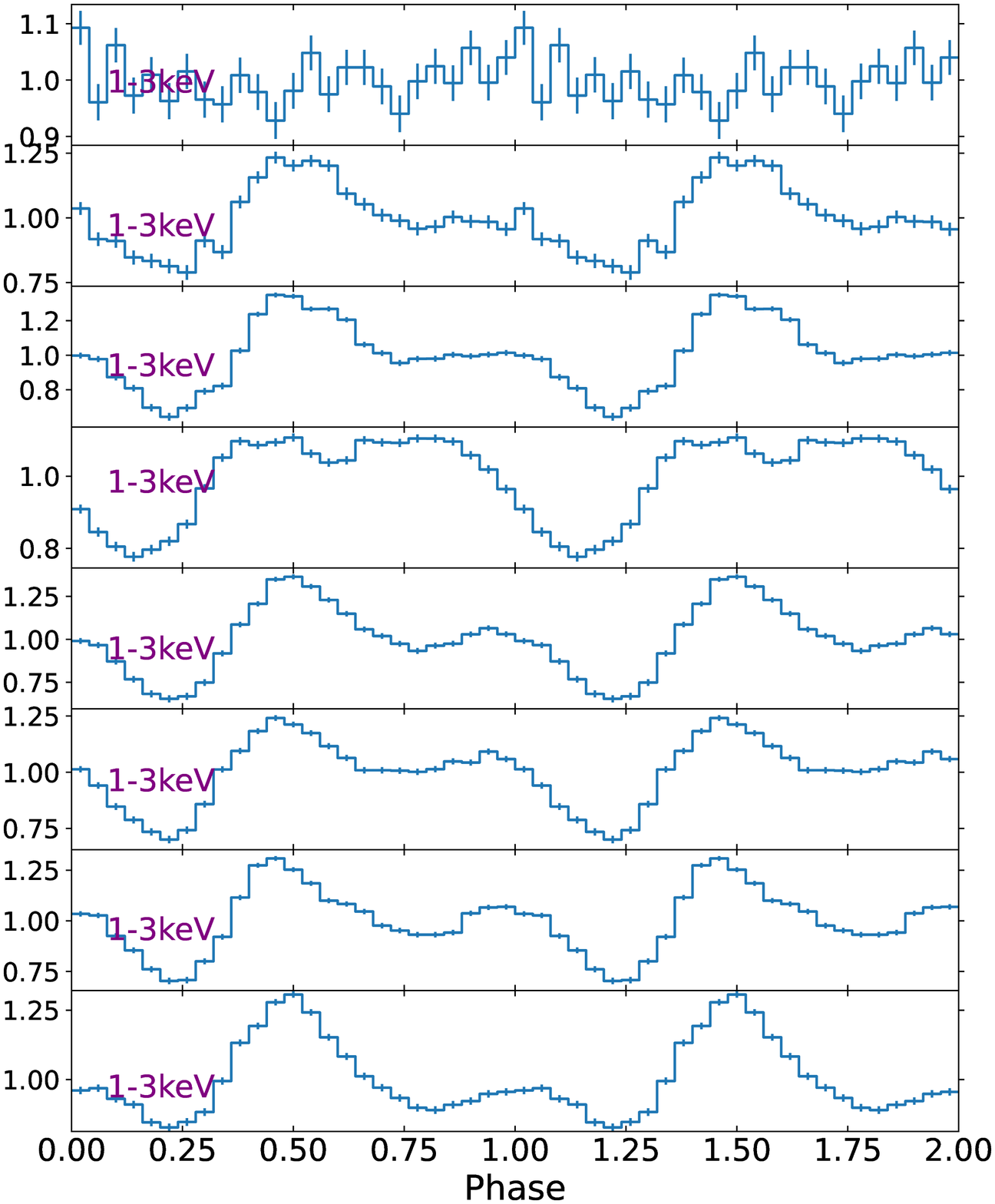}
    \includegraphics[width=.3\textwidth]{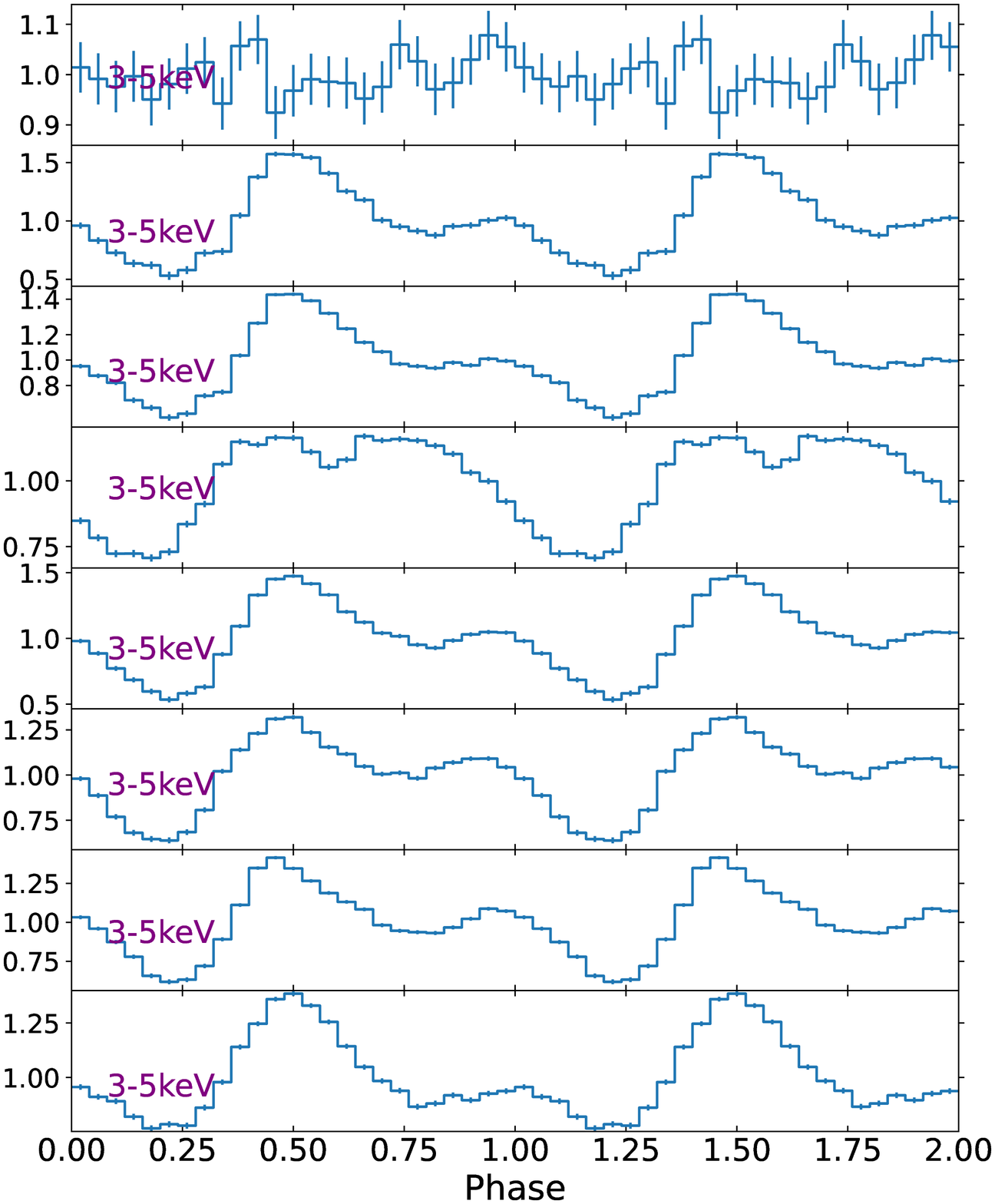}
    \includegraphics[width=.3\textwidth]{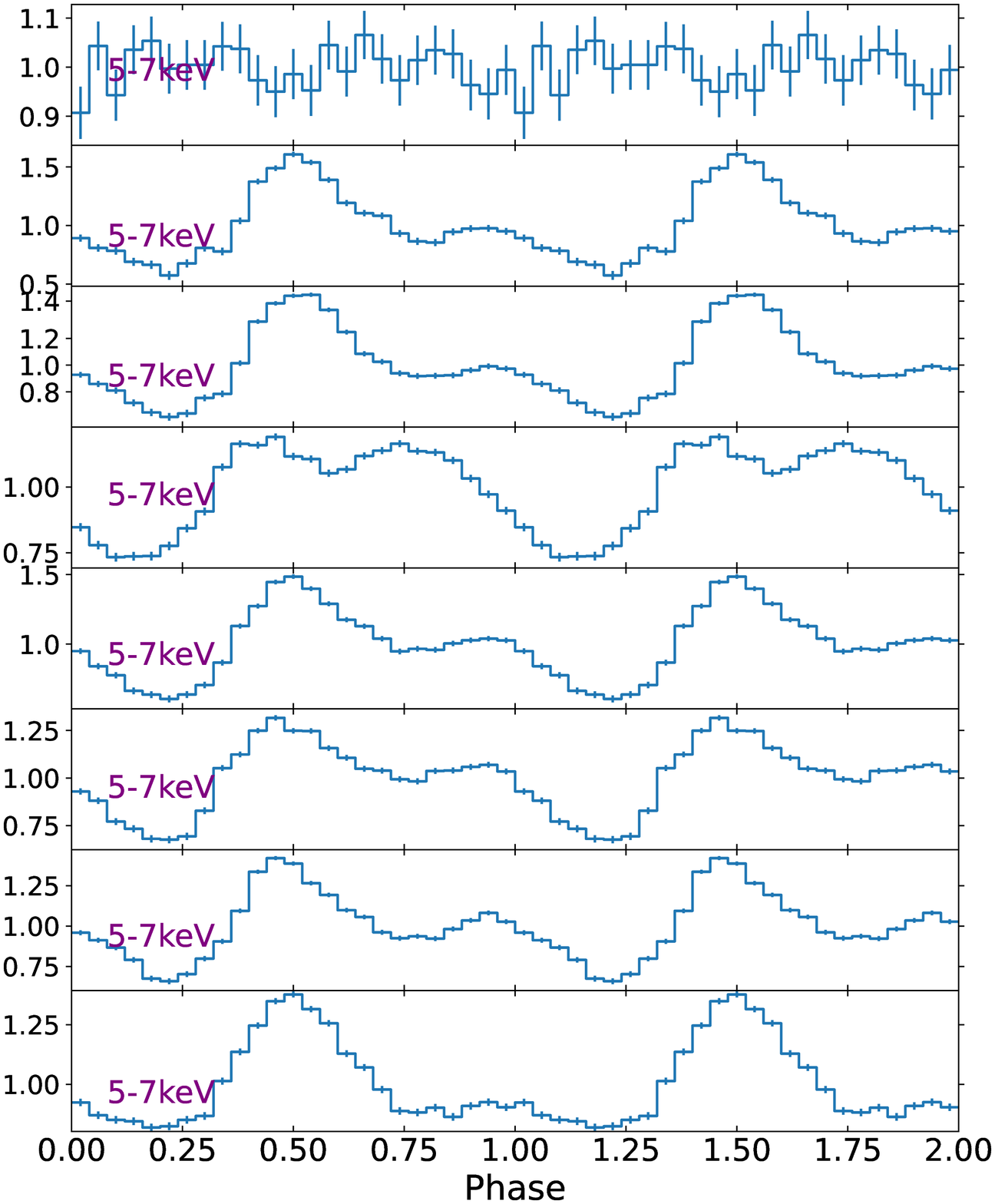}
    \includegraphics[width=.3\textwidth]{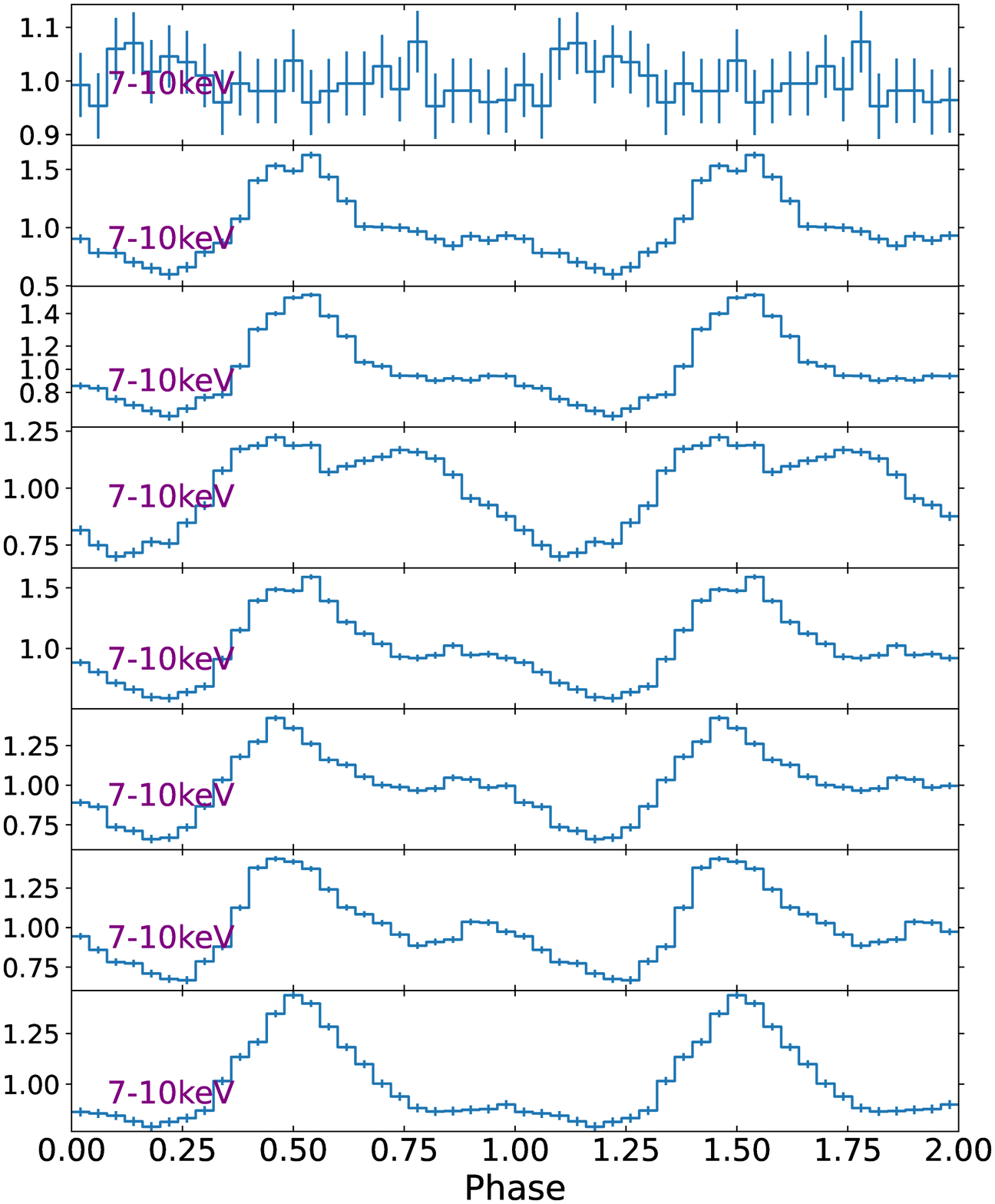}
    \includegraphics[width=.3\textwidth]{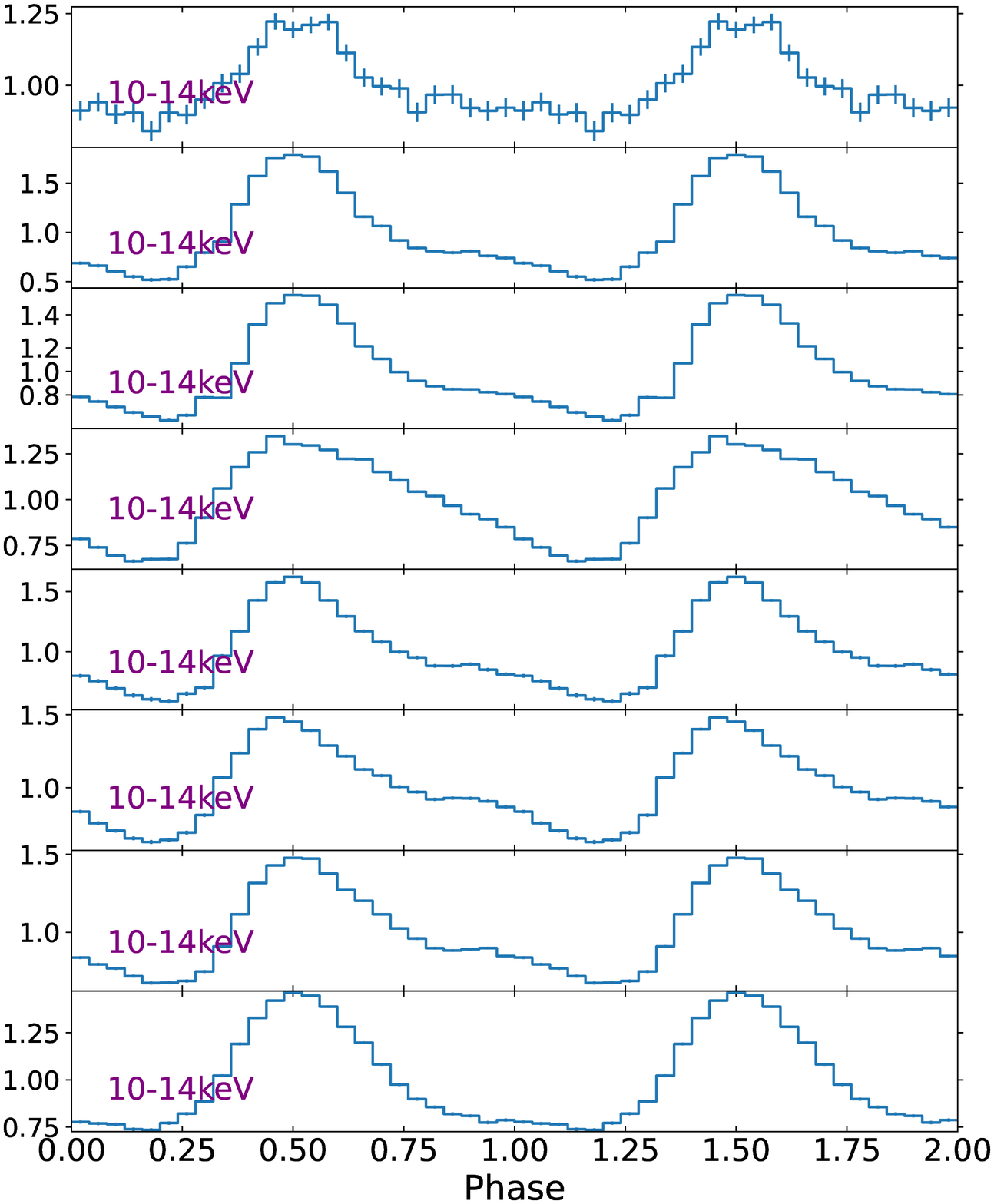}
    \includegraphics[width=.3\textwidth]{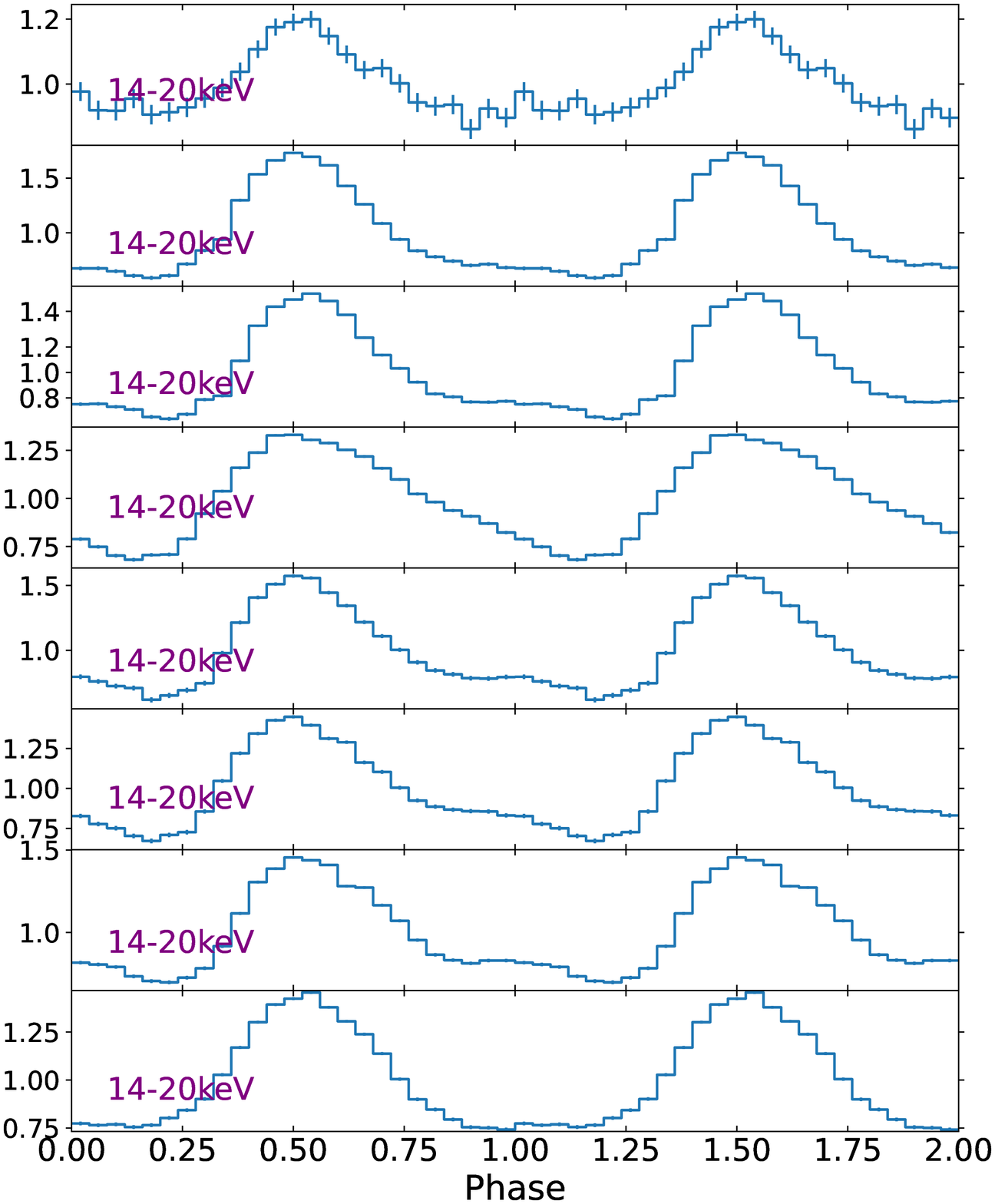}
    \includegraphics[width=.3\textwidth]{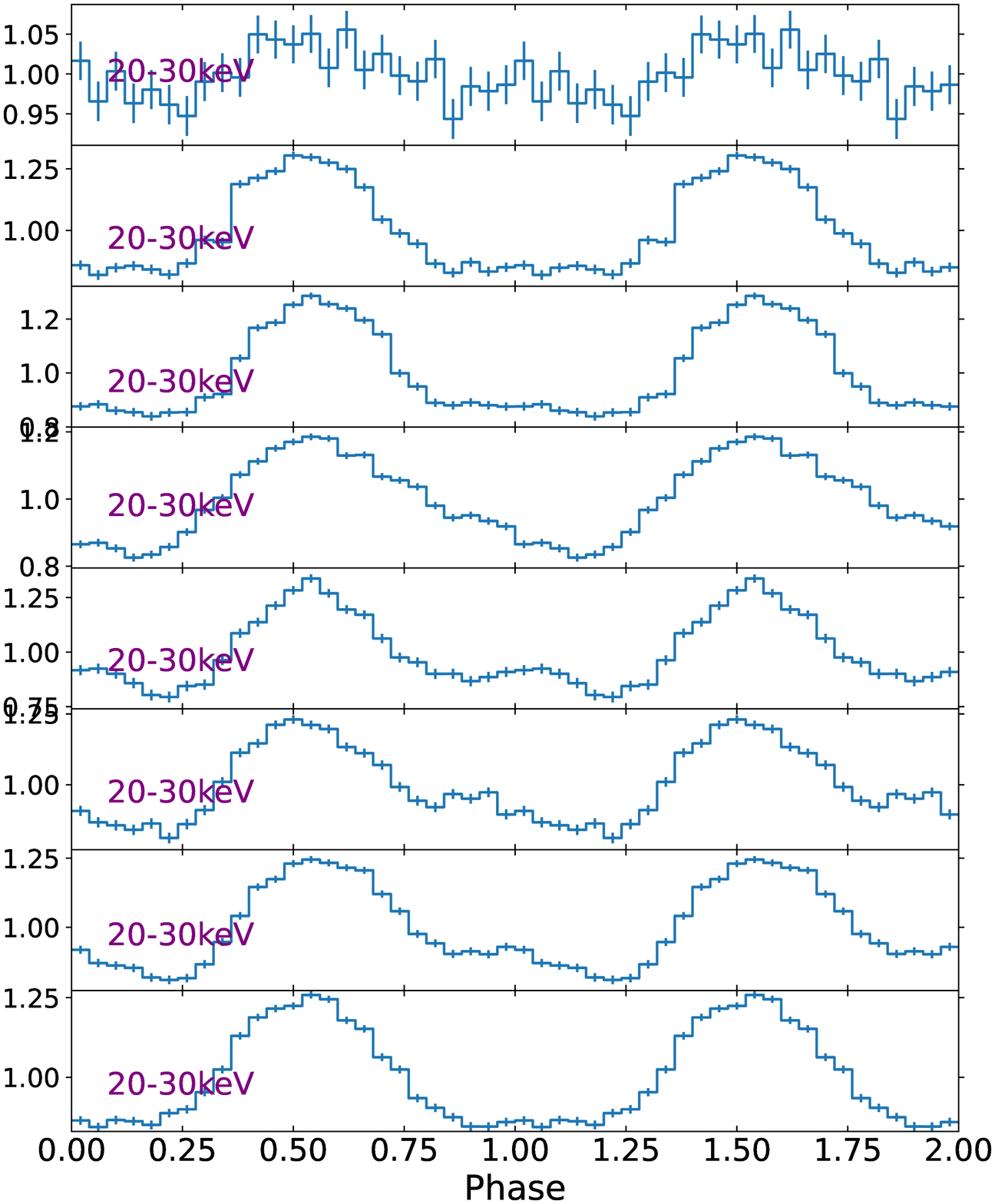}
    \includegraphics[width=.3\textwidth]{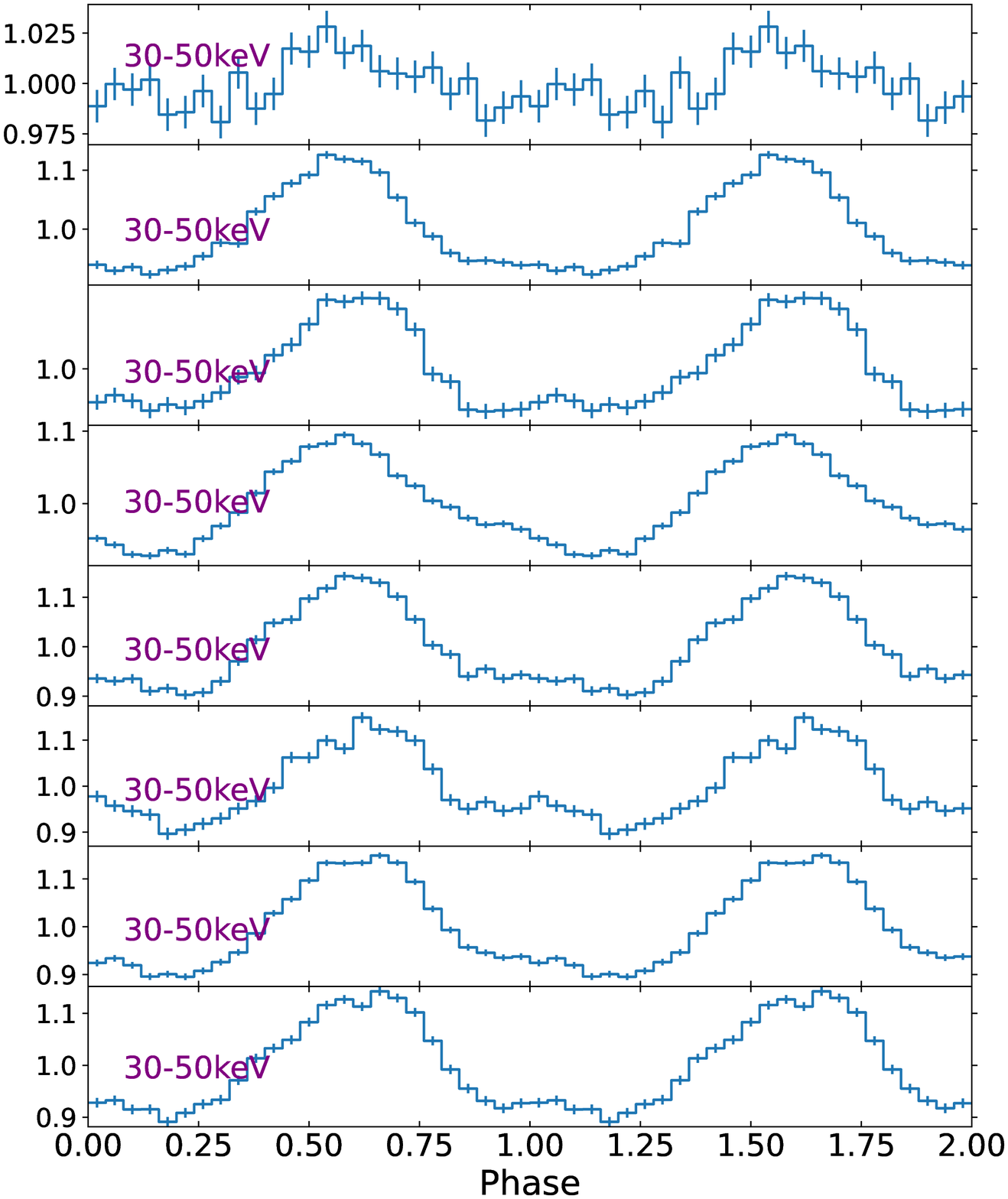}
    \includegraphics[width=.3\textwidth]{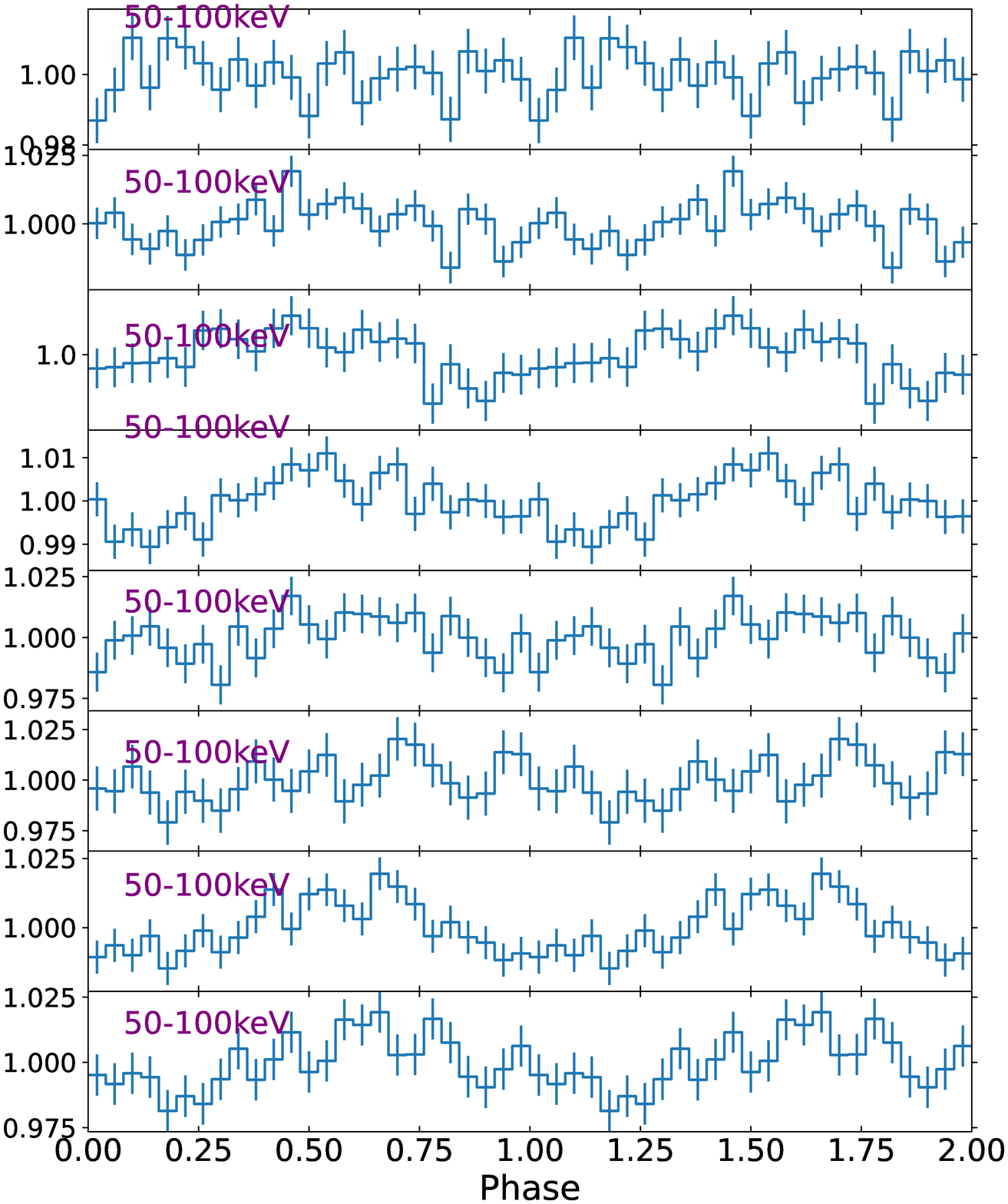}
    \caption{The evolution of pulse profiles of Cen X-3 on February 04, 2018 over nine energy bands of 1–3 keV, 3–5 keV, 5–7 keV, 7–10 keV, 10-14 keV, 14-20 keV, 20-30 keV, 30-50 keV and 50-100 keV. For each panel, from top to bottom, the profiles for eight time segments during the observation are also plotted for the comparison.}
    \label{fig:pulse_time}
\end{figure*}

\begin{figure*}
    \centering
    \includegraphics[width=.3\textwidth]{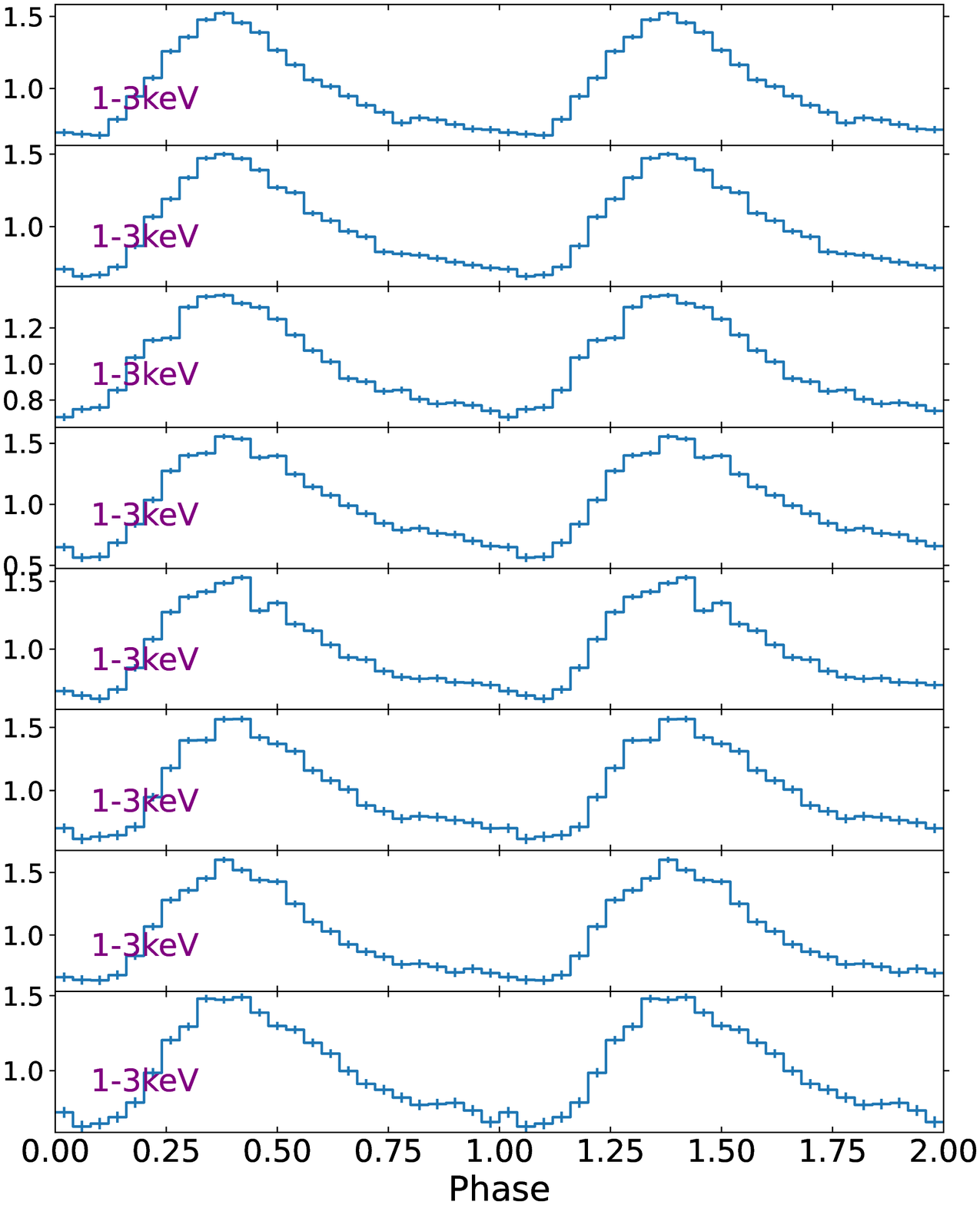}
    \includegraphics[width=.3\textwidth]{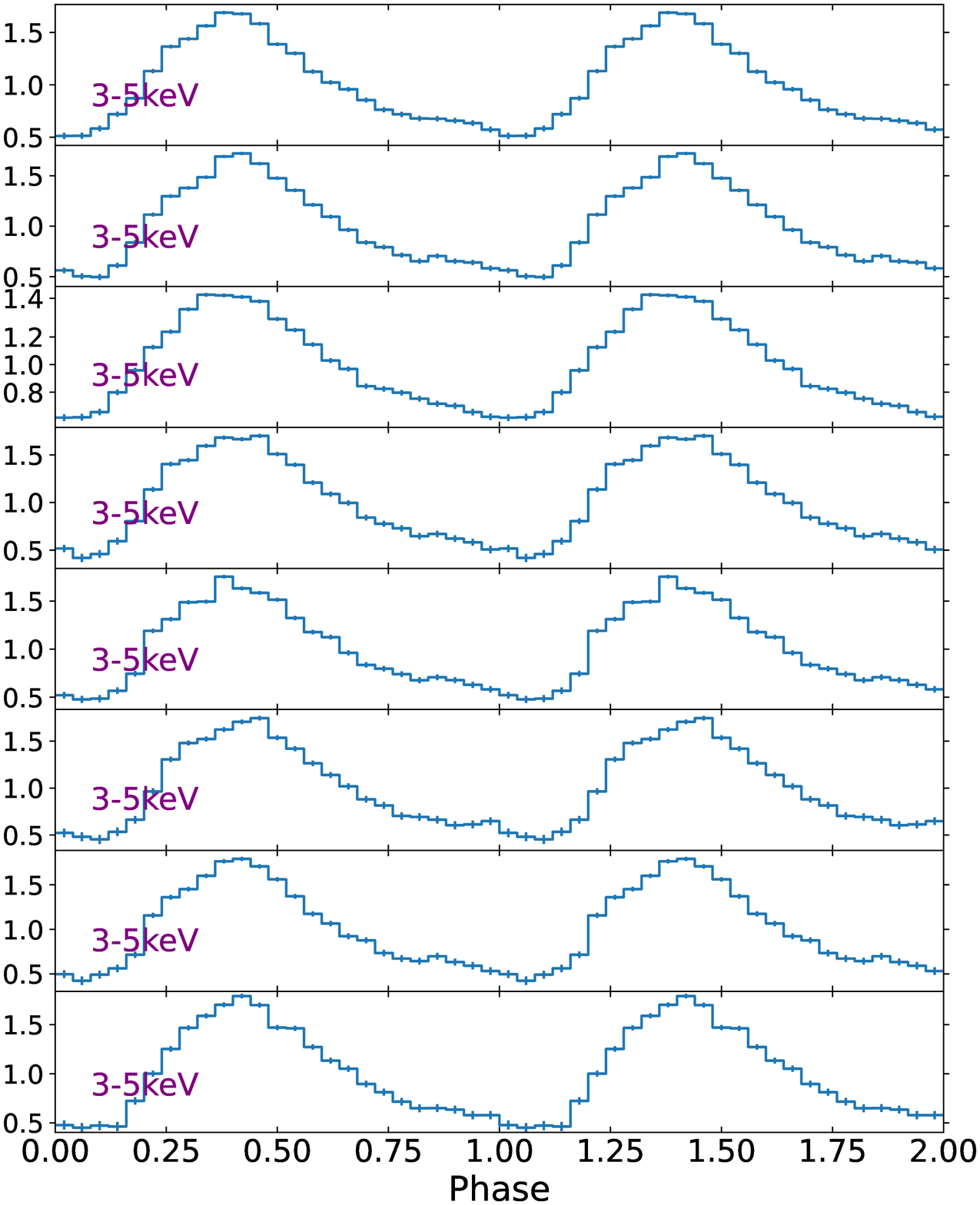}
    \includegraphics[width=.3\textwidth]{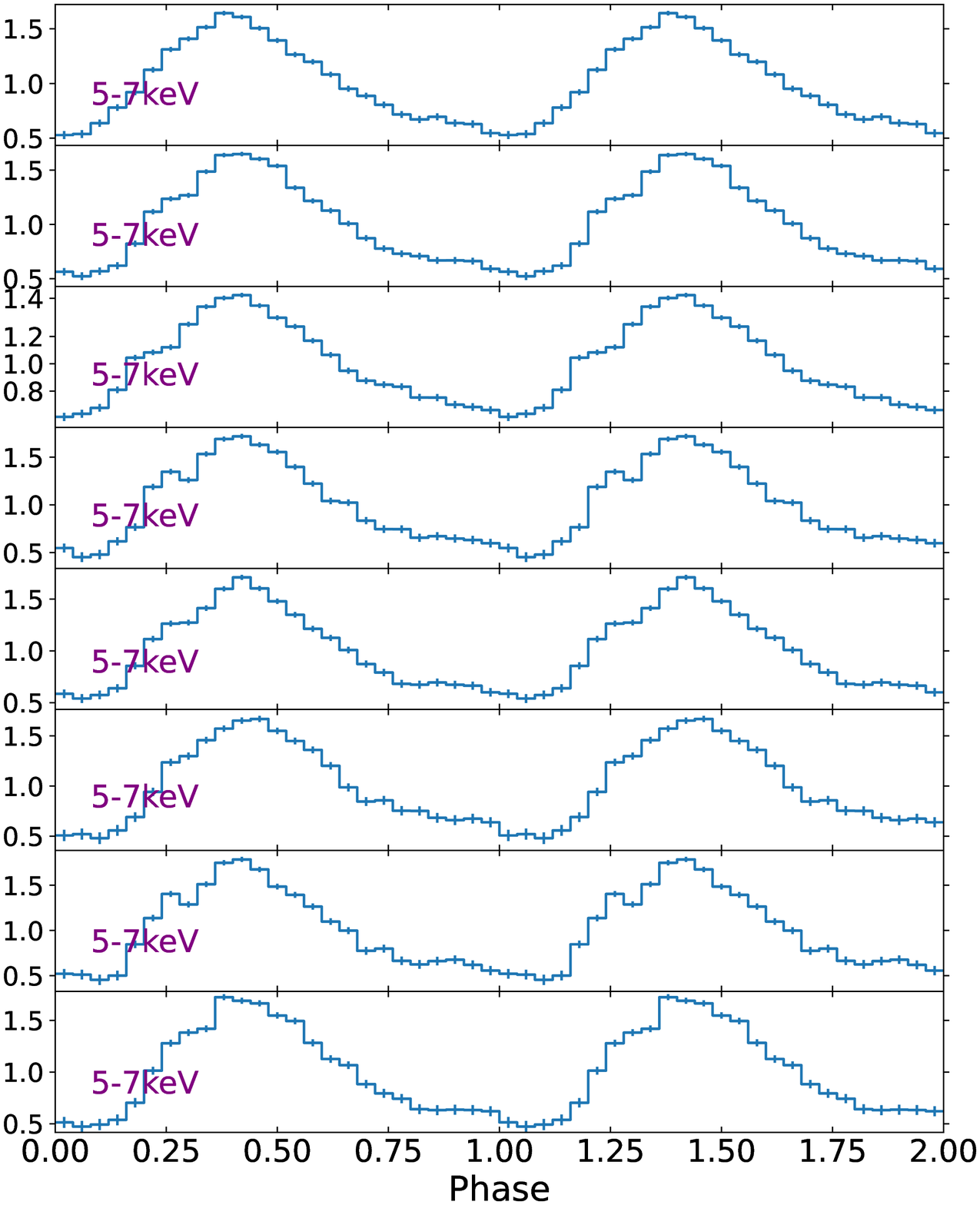}
    \includegraphics[width=.3\textwidth]{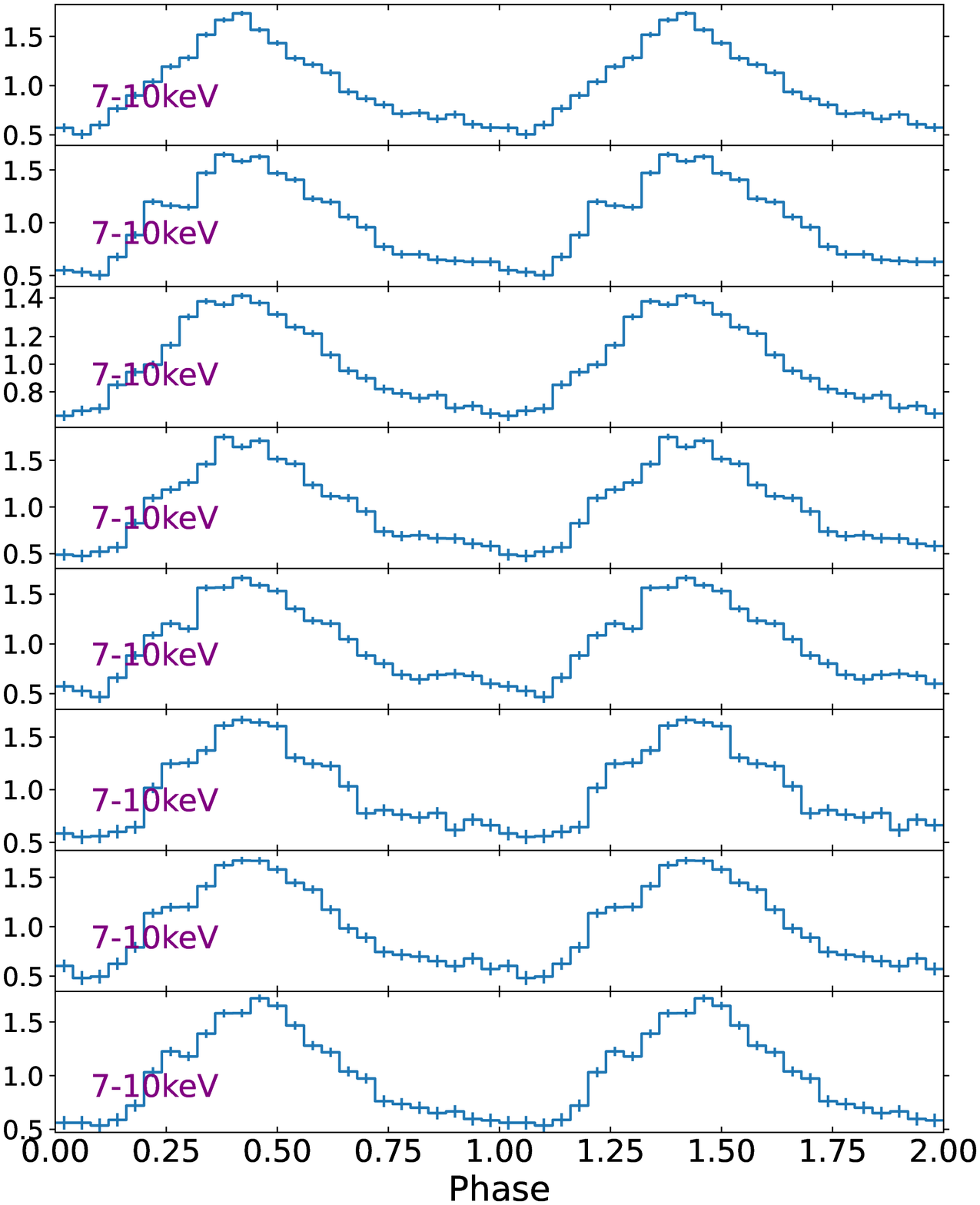}
    \includegraphics[width=.3\textwidth]{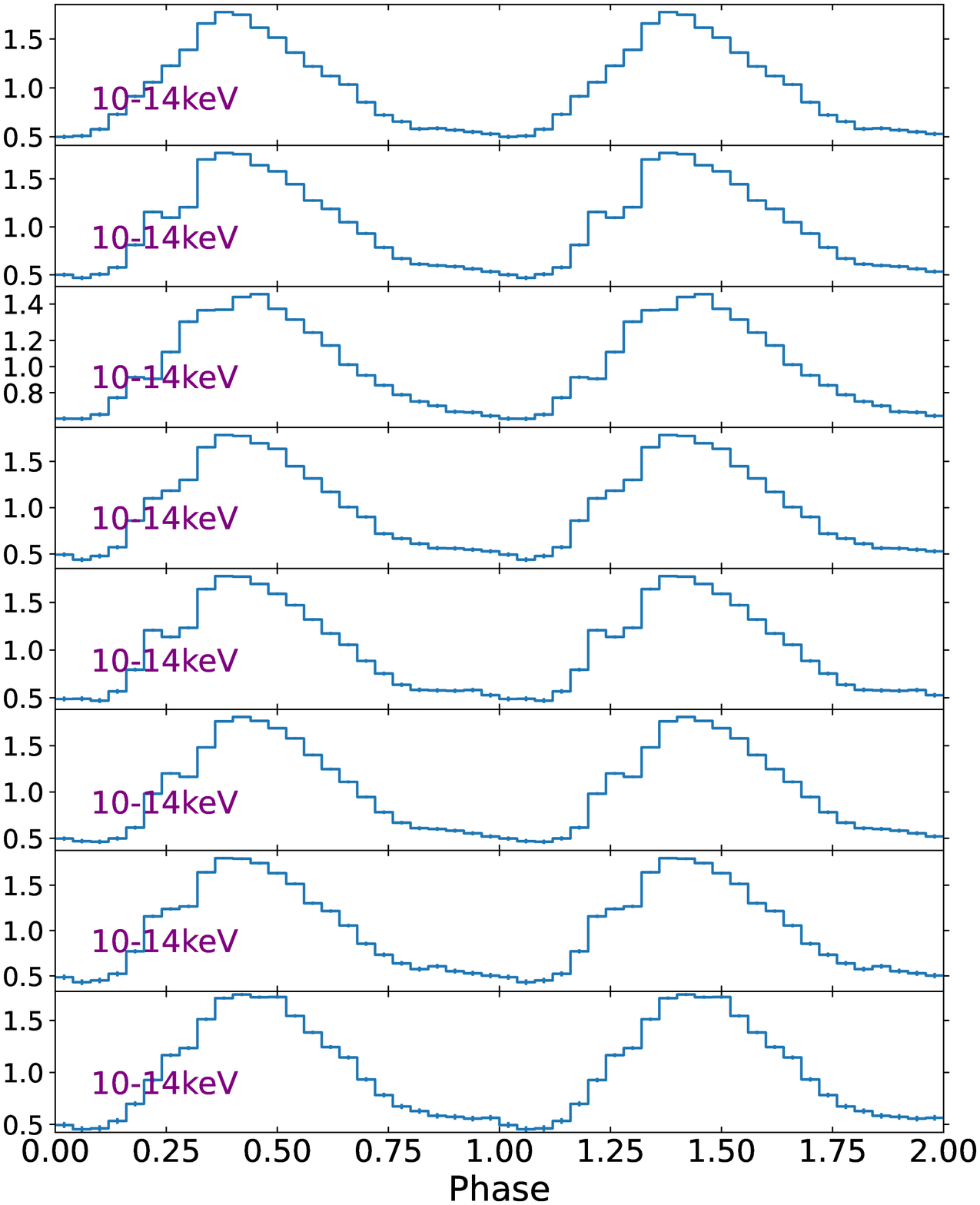}
    \includegraphics[width=.3\textwidth]{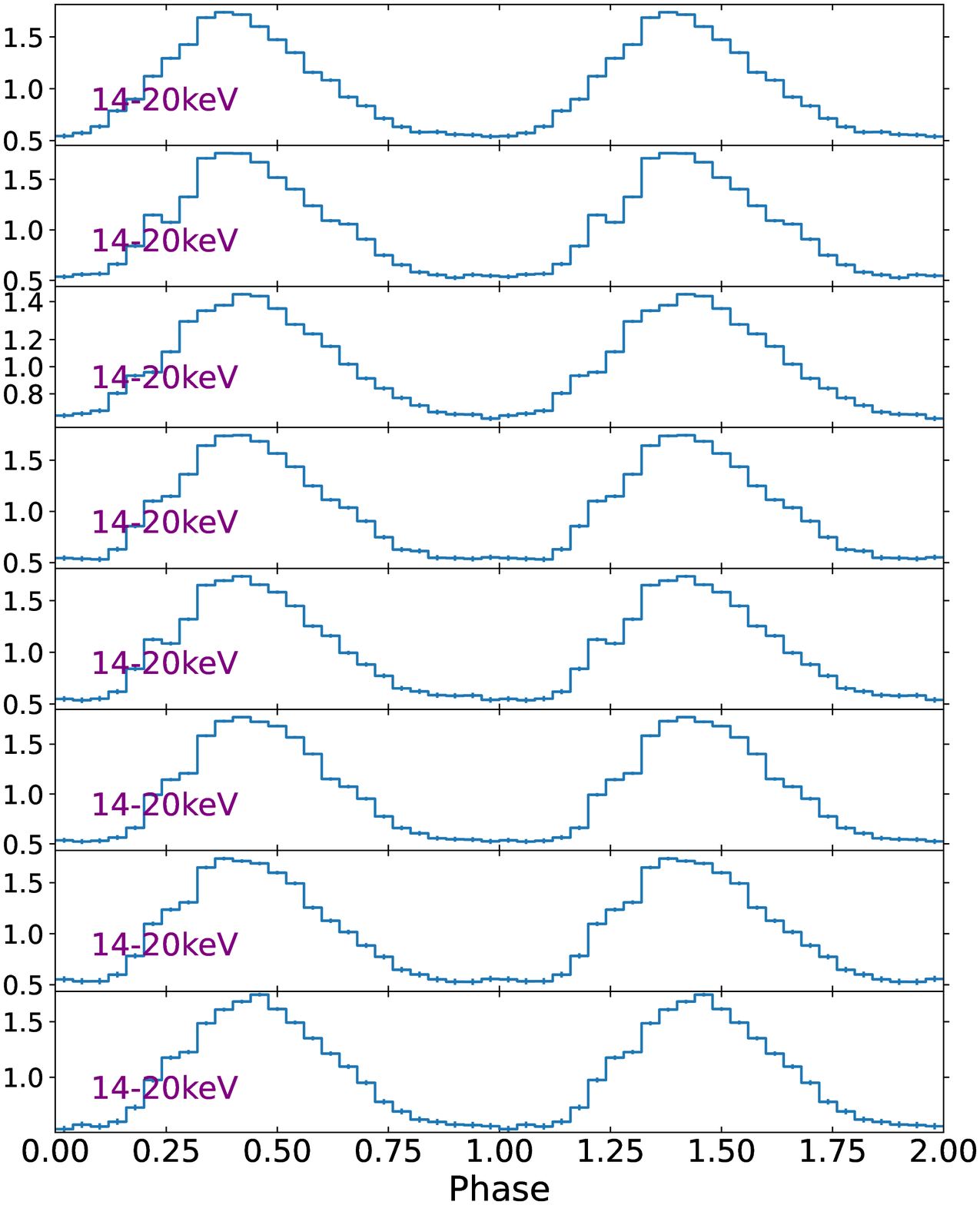}
    \includegraphics[width=.3\textwidth]{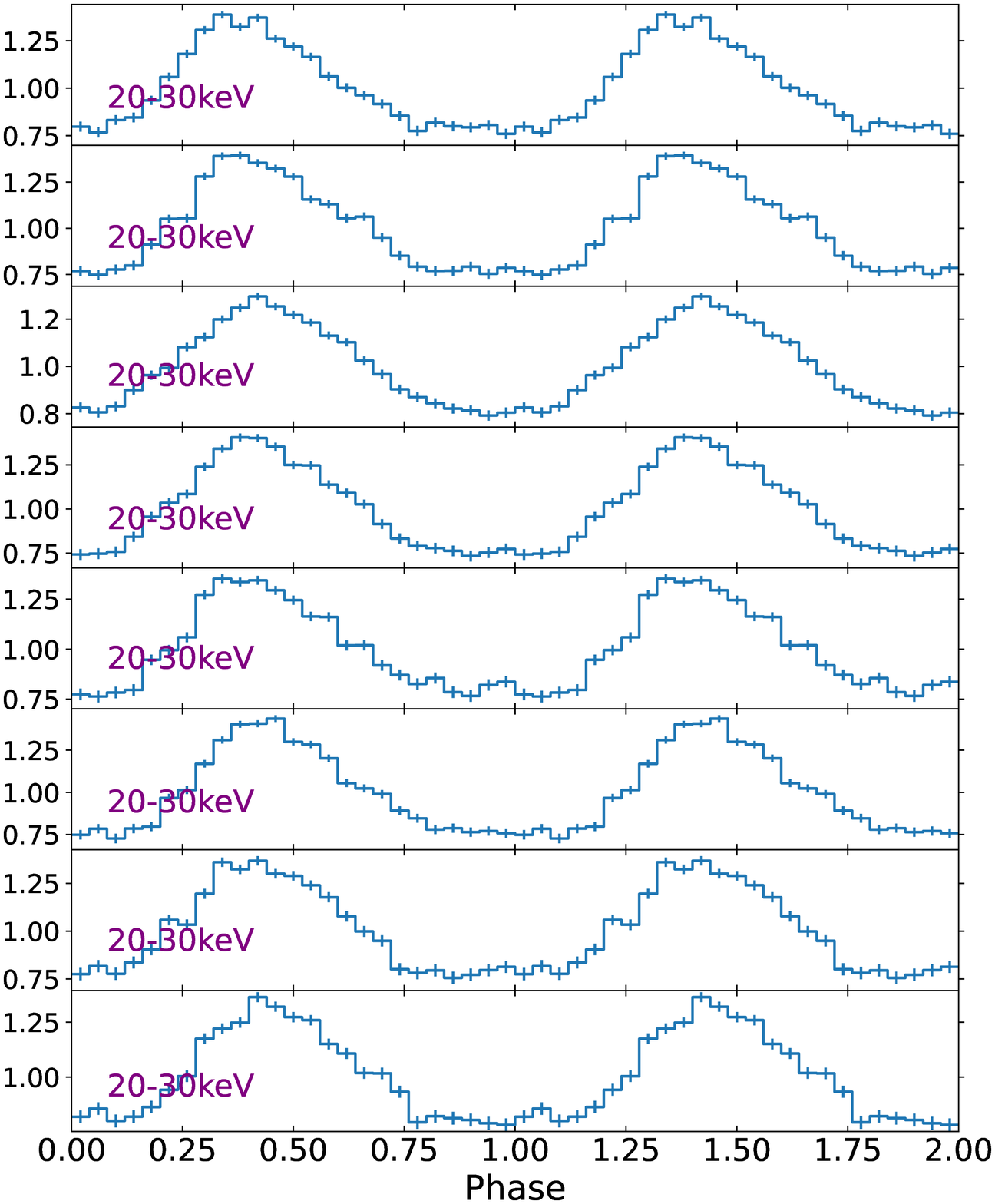}
    \includegraphics[width=.3\textwidth]{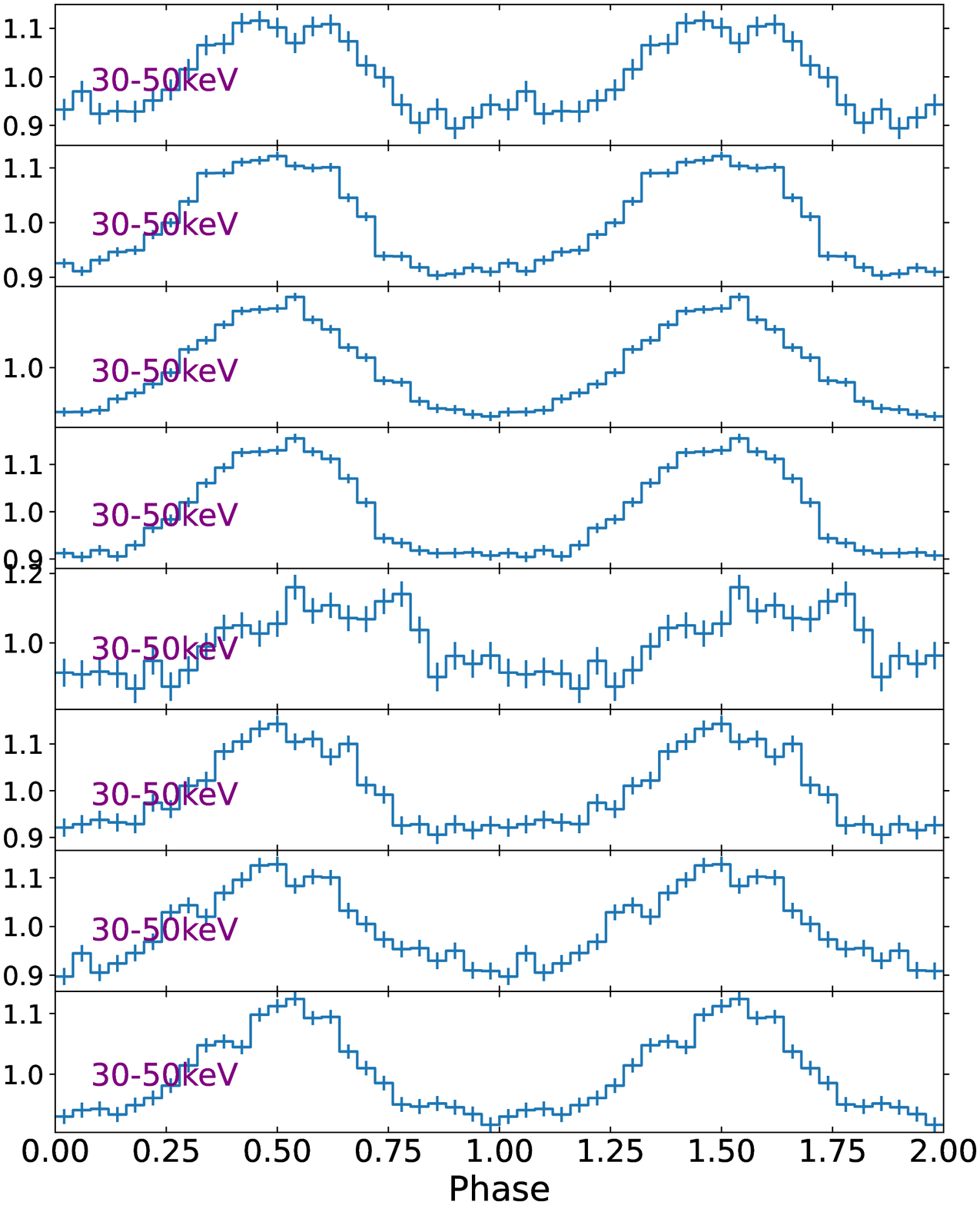}
    \includegraphics[width=.3\textwidth]{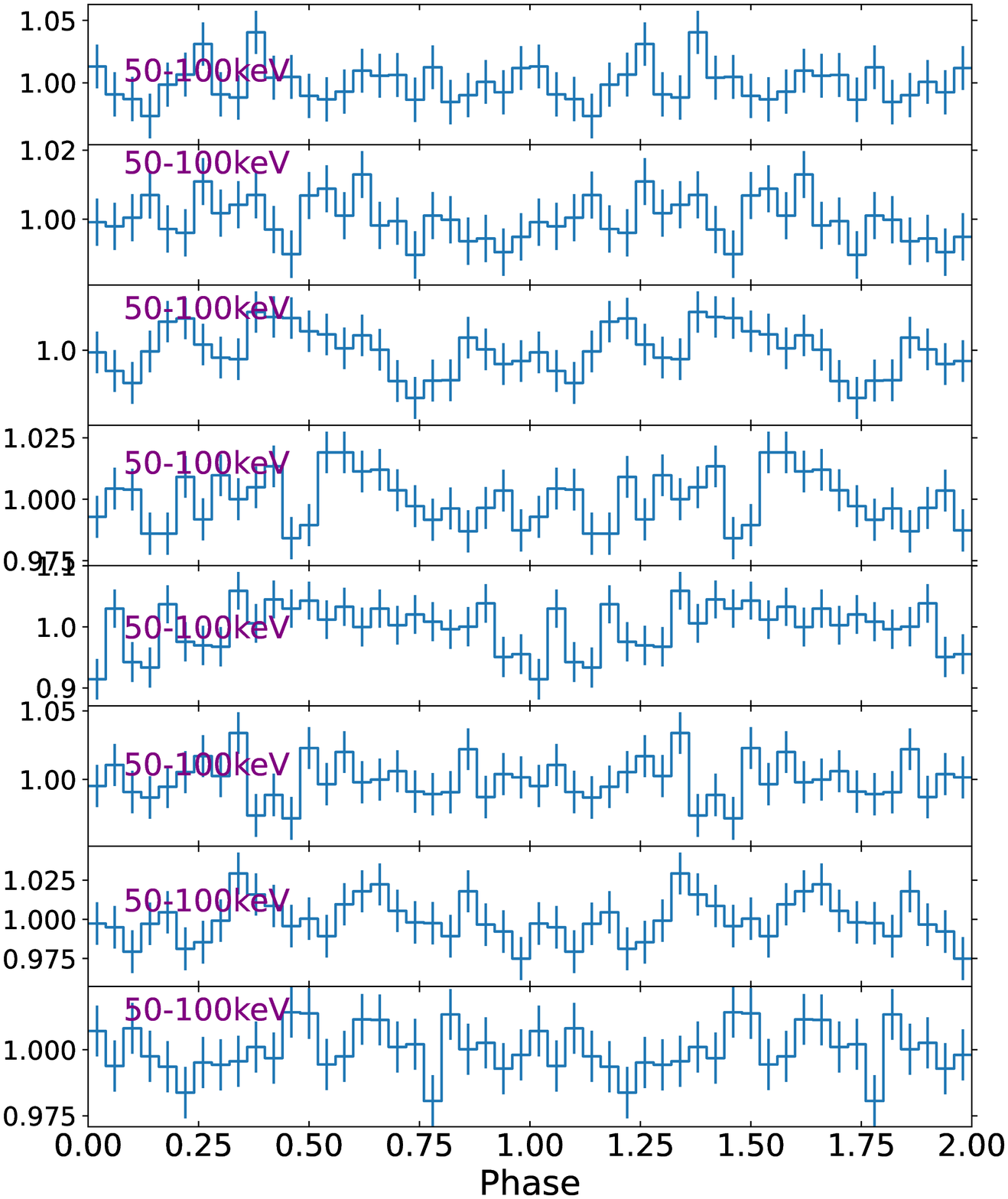}
    \caption{The evolution of pulse profiles of Cen X-3 on January 05, 2020 over nine energy bands of 1–3 keV, 3–5 keV, 5–7 keV, 7–10 keV, 10-14 keV, 14-20 keV, 20-30 keV, 30-50 keV and 50-100 keV. For each panel, from top to bottom, the profiles for eight time segments during the observation are also plotted for the comparison.}
    \label{fig:pulse_time02}
\end{figure*}

We investigated the energy-resolved pulse profiles into different energy bands of 1–3 keV, 3–5 keV, 5–7 keV, and 7–10 keV for LE, 10-14 keV, 14-20 keV, and 20-30 keV for ME, and 30-50 keV and 50-100 keV for HE at a phase resolution of 30 phase bins/period. Figure\ \ref{fig:pulse} also shows the energy-dependent pulse profiles over different energy bands for the ExpID P010131100809 generated by folding with the best period. We detected a double-peak profile at a soft energy range of $1-10$ keV with a prominent second peak. However, the secondary peak began to fade away at higher energies and the pulse shape became to be dominated by a single peak. A similar energy-dependence pulse profile with a significant secondary peak at lower energies are also reported by previous results \citep{1992ApJ...396..147N,2000ApJ...530..429B,2008ApJ...675.1487S,2021JApA...42...58S}. We also obtained the hardness ratio for 3-5 keV/1-3 keV. This ratio indicates that the secondary peak is softer than the main peak, which is consistent with the results reported by \cite{2000ApJ...530..429B}. \cite{2008ApJ...675.1487S} showed the peaks in the different hardness ratios with RXTE data and explain the beam emission in lower X-ray energy. But our results indicate that the emission region of higher energy is more pencil-like. \cite{2007ApJ...654..435B} assumed the accretion column model, and thought that the X-ray emission from the top of the column show a pencil beam, and that emission from the wall of the column is like a fan beam. The energy-resolved pulse profile with double-peaked at lower energies and single-peaked at higher energies may indicate the emission is a mixture of the pencil beam and fan beam patterns as suggested by \cite{Tamba_2023}.

To study the evolution of the pulse profile with time, we used two observations on February 04, 2018 (divided into 8 segments in approximately one orbit) and January 05, 2020. Each observation covered near one orbit, and we also divided it into eight time segments (see Table \ref{tab:ExpIDs}) to derive the pulse profiles of each time segment. Figure\ \ref{fig:pulse_time} shows the pulse profile obtained at different times in 2018 over nine energy ranges. As we can see, the pulse profile has a prominent double peak at the band of 1-3 keV (the start time of MJD 58153.4, we ignored the first panel due to lower flux). As time goes on, the secondary pulse increases, and the two peaks seem to be comparable at MJD 58154.2 (the fourth panel of Figure\ \ref{fig:pulse_time}). After the MJD 58154.2, the secondary peak began to decrease. The profile of 1-10 keV evolves with the same behavior. As the energy increases, the pulse profile (10-100 keV) slowly evolves into a broad single peak.  However, these pulse profiles in 2020 (the start time of MJD 58853.1) only has a single-peaked profile at all energy ranges as seen in Figure\ \ref{fig:pulse_time02}, and do not show obvious changes in pulse profiles at low energies with the time.

\begin{figure}
    \centering
    \includegraphics[width=.5\textwidth]{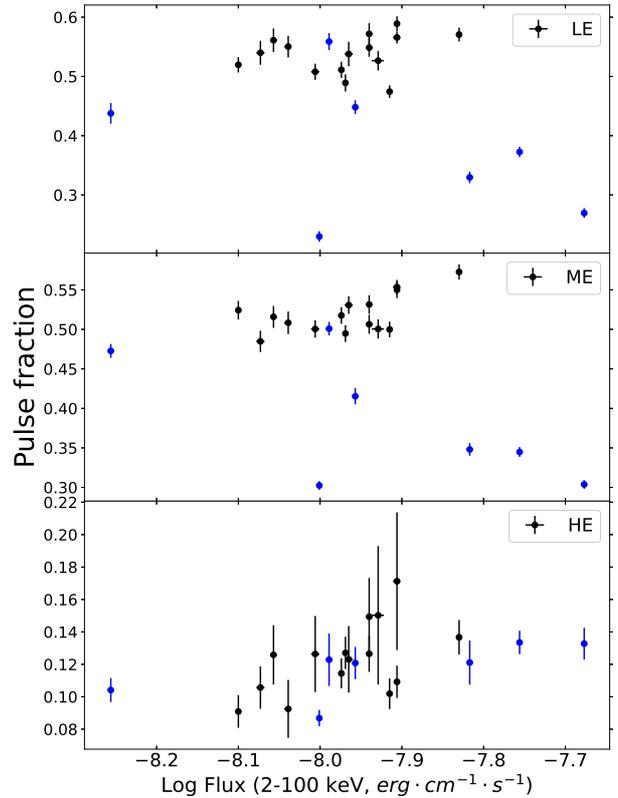}
    \caption{The pulse fractions of Cen X-3 (i.e., PF=($I_{max}$-$I_{min}$)/($I_{max}$+$I_{min}$).) over three energy bands (LE:2-10 keV; ME:10-30 keV; HE:30-50 keV) versus X-ray flux. The blue points represent the data from 2018 and the black points from 2020.}
    \label{fig:pulse_fraction}
\end{figure}

The pulse fractions of the neutron star X-ray binary in three energy bands are plotted against an un-absorbed X-ray flux in Figure\ \ref{fig:pulse_fraction}. The flux was estimated by fitting the spectrum in the 2–100 keV with a model of $TBabs*pow*highecut$ along with the cyclotron absorption lines and Fe line in XSPEC \citep{1996ASPC..101...17A}. The \textsl{cflux} in XSPEC was used to calculate an unabsorbed flux. The averaged pulsed fractions are about 0.55, 0.50, and 0.13 in the three energy bands (2-10 keV, 10-30 keV and 30 -50 keV), respectively. In Figure \ \ref{fig:pulse_fraction}, the black data points show the pulse fraction observed in 2020 versus the flux, the positive relation appears in different energy bands, specially it is stronger in the HE band, which would be consistent with the theoretical expectation \cite{1975A&A....42..311B,1976MNRAS.175..395B}. While for 2018 observations, the data points (blue) of the LE and ME bands have a large variation even at a similar flux state, but for the HE band, the pulse fraction seems to have a weak positive relation with the flux. Then we didn't find a correlation between pulse fraction and flux as expected by varying accretion rates in 2018, and the pulse profiles in soft X-ray bands even evolved with time. And the pulse fraction in high energy has a lower pulse fraction, which may be due to the main peak being slightly smeared for the low count rates in high energy bands.

Based on the theory proposed by \cite{1975A&A....42..311B,1976MNRAS.175..395B}, the variations of pulse profiles at low energies ($<$ 10 keV) may be related to the varying accretion rate, which would also result in the correlation between the pulse fraction and accretion rate. Though the flux may not represent the true accretion rate due to a varying degree of obscuration by a precessing warped accretion disc in Cen X-3 \citep{2008MNRAS.387..439R}, pulse fraction versus flux in 2018 show no relation in low energies, and different fractions in similar flux, which contradicts with the model prediction. Since the flux of the source is in the similar flux level in 2018 and 2020, it is possible that the emission patterns have changed in the time between 2018 and 2020. The complicated profiles in 2018 indicates that the emission from the polar caps is still a combination of a pencil beam and a fan beam \citep{Tamba_2023}, whereas a simple single peak in 2020 may imply the emission is dominated by a pencil beam. There also exist other possibilities, such as the different shapes of hot spots (i.e., a ring-like hollow or a filled circle) on the surface of the neutron star, even the emission contributed by the reflection component \citep{2013ApJ...777..115P}, then the changes in pulse profiles at low energies could be ascribed to the contributions from the different components.

\subsection{The spin and orbital parameters}

Traditionally, one can estimate the pulse period and orbital parameters by measuring the pulse arrival time from a neutron star in binaries. However, the pulse arrival time is likely to be affected by the pulse profile variations. So in our work, we have used the pulse period evolution (e.g., \citealt{2010MNRAS.406.2663R,2012MNRAS.423.2854L,2021MNRAS.503.6045D}) instead of using pulse arrival time to determine the orbital parameters. 
The observed pulse periods $p_{\mathrm{obs}}$ are modified by the binary orbital motion of the system. To obtain the intrinsic pulse period $p_{0}$ of the neutron star from the observed values, we must correct this Doppler modulation based on the following relation:
\begin{equation}
p_{\mathrm{obs}}\left({\mathrm{t}}\right)=p_{0}\left(1+\frac{v\left({\mathrm{t}}\right)}{c}\right).
\end{equation}
We assume a circular orbit (also see \citealt{2000ApJ...530..429B,Tamba_2023}) since the eccentricity is nearly zero (< 0.0001, \citealt{2010MNRAS.401.1532R}), and the radial velocity v(t) along the line-of-sight can be characterized by a sinusoidal function:
\begin{equation}
v\mathrm{(t)}= A \sin [2\pi \phi],
\end{equation}
where $\phi$ is the orbital phase of the binary, and $\phi =0 $ corresponding to the mid-eclipse $T_{\rm ecl}$. The orbital phase can be replaced by $\phi = \frac{(t-T_{\rm ecl})}{P_{\rm orb}}$. The amplitude A of the radial velocity v(t) can be expressed with the following orbital parameters
\begin{equation}
A=\frac{2 \pi a_{x} \sin i}{P_{\rm orb}}.
\end{equation} 
where $P_{\rm orb}$ denotes the orbital period and $a_{x} \sin i$ is the projected semi-major axis. Combining Equations (2) and (3), one can get
\begin{equation}
p_{\mathrm{obs}}\left({\mathrm{t}}\right)=p_{0}\left(1+ \frac{2 \pi a_{x} \sin i}{P_{\rm orb}} \sin \left[\frac{2 \pi\left(t-T_{\rm ecl}\right)}{P_{\rm orb}}\right]\right).
\end{equation} 

We fitted the observed spin periods with the function above. The fitting is based on the least-squares fitting and the errors on the spin period are weighted when fitting the orbital parameters. And the errors on spin period measurement are multiplied by a constant factor to get a reduced $\chi^2$ of $\sim$1. The estimated error for orbital parameters is computed from the covariance matrix. 

\section{Results and Discussion} \label{sec:discussion}

To test the performance, we first performed a fitting of the light curves for the 2018 observations and found the orbital period to be $2.087\pm 0.001$ d, which is still consistent with the value derived by \cite{2015AA...577A.130F}. However, the obtained orbital period only based on the Insight-HXMT data has a large error bar and the precision is not so good compared with the previous observations, maybe be not a best choice to perform the analysis of the orbital parameters, therefore, we fix the orbital period $P_{\mathrm{orb}}$ with a value of 2.08704106 d at MJD 50506 reported by \cite{2015AA...577A.130F} and considered its derivative in the fitting.

\begin{figure}
\centering
\includegraphics[width=0.5\textwidth]{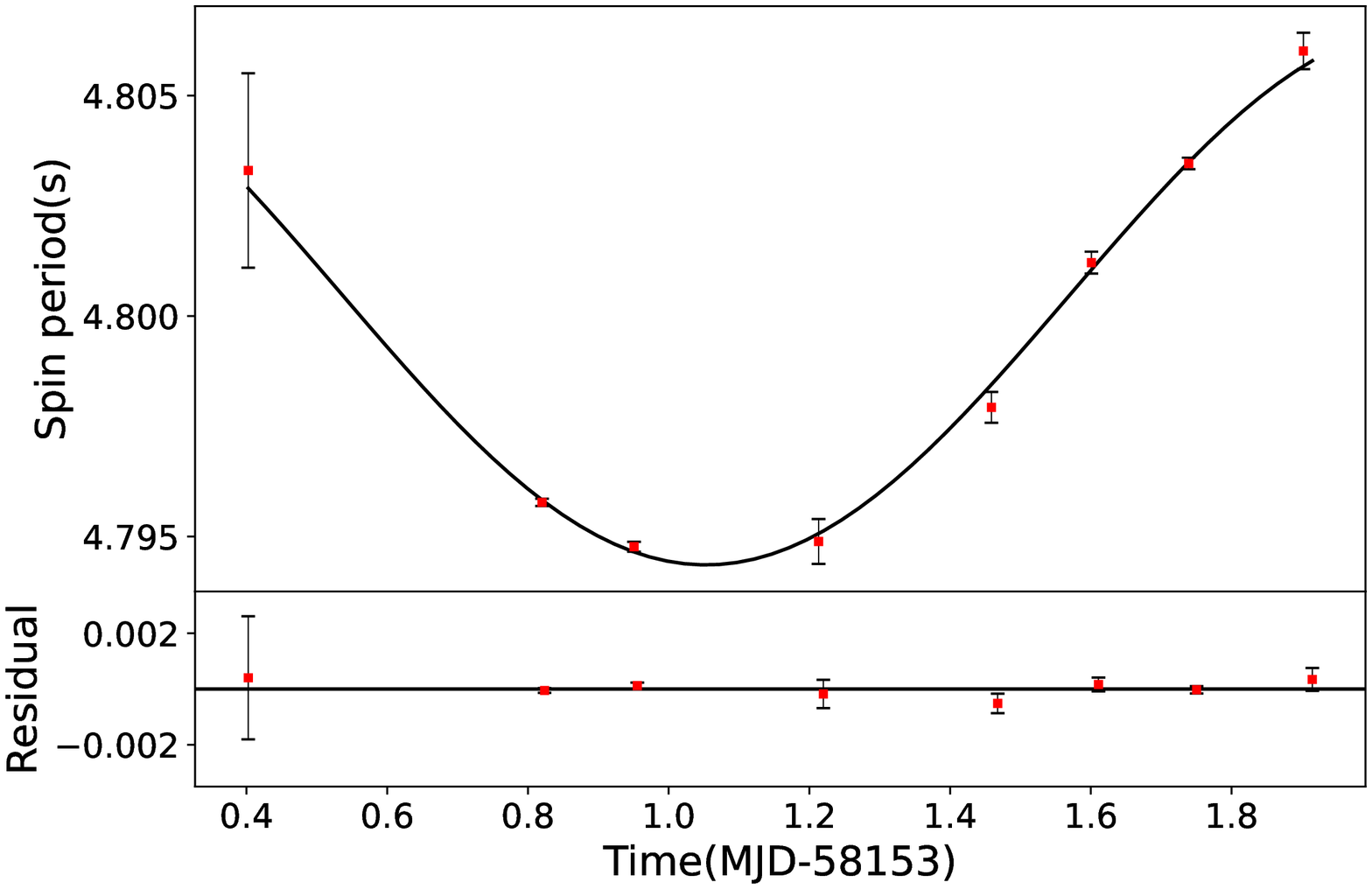}
\includegraphics[width=0.5\textwidth]{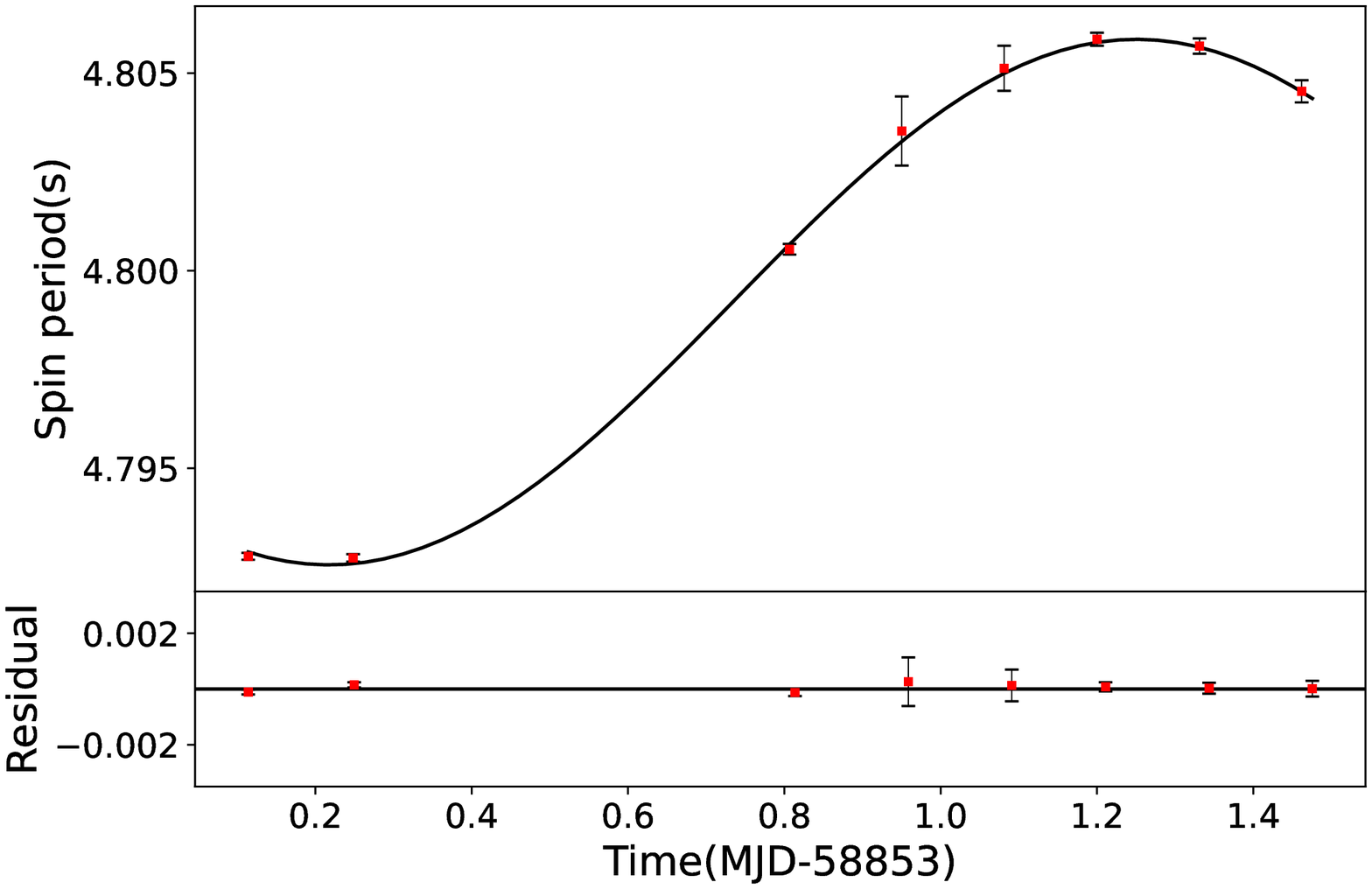}
\includegraphics[width=0.5\textwidth]{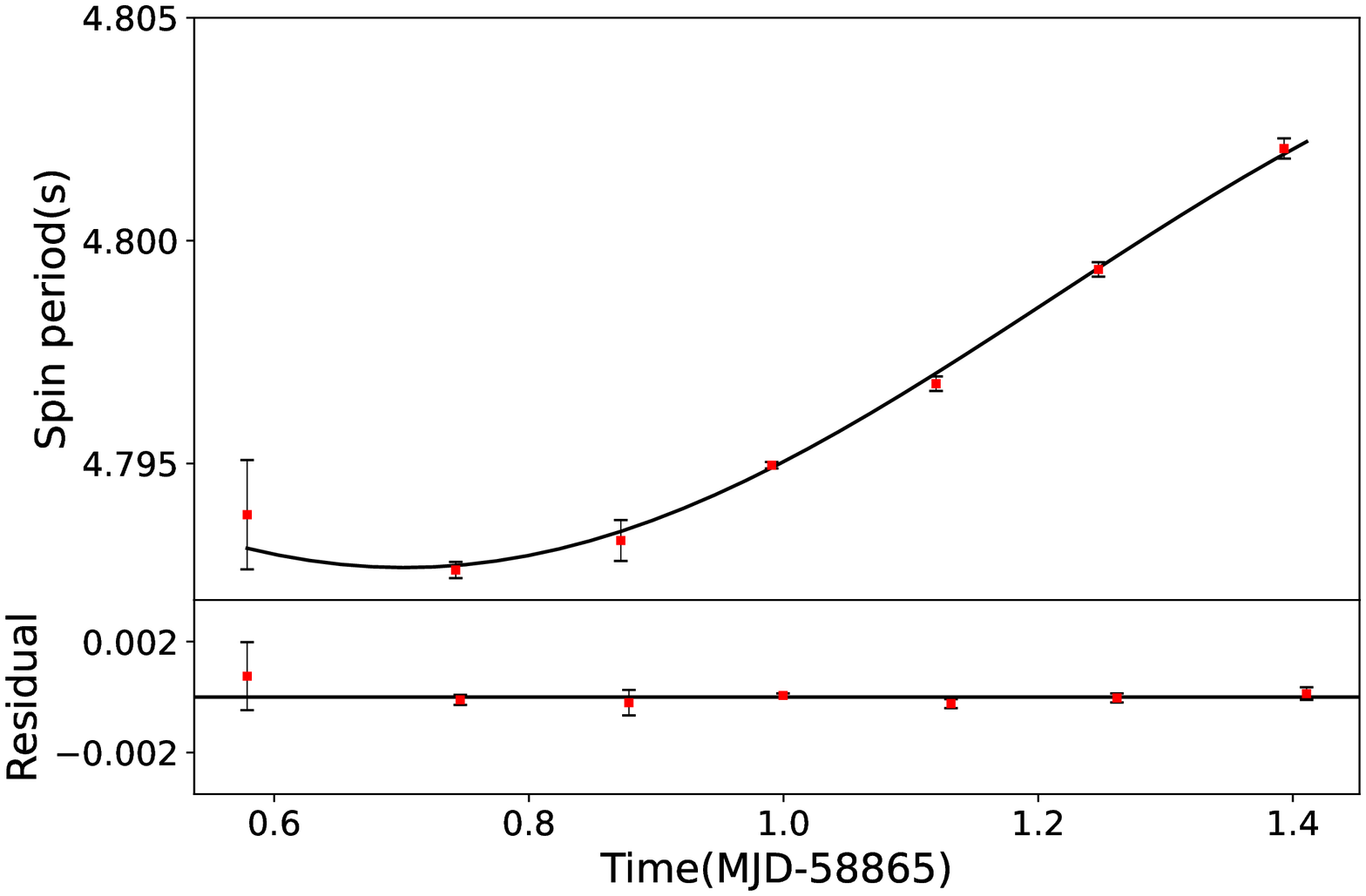}

\caption{The available observed spin period values of Cen X-3 during the 2018 and 2020 observations 
based on Insight-HXMT. The black solid curve represents the best fitting model due to the Doppler motion of the binary. Residuals are shown in the lower panel.}
\label{fig:orbit}
\end{figure}

Figure\ \ref{fig:orbit} shows the best-fitting plots from the three available observations in 2018 and 2020, and a clear sinusoidal variation is present. The fluctuations of residuals for Cen X-3 may be due to a very small eccentricity but we are unable to constrain it accurately. The fitting results including the intrinsic spin period, the eclipse epoch, and the projected semi-major axis for the observations in 2018 and 2020 are listed in Table\ \ref{tab:orbital}. The results for 2018 and 2020 are generally consistent with each other under certainties. Taking a rough estimation from MJD 58153 to MJD 58853, we can obtain the spin derivative of $\sim -0.5$ ms per year, 
which is consistent with a spin-up trend for the disc accretion (sometimes with some wavy fluctuations), as revealed in long-term monitoring (see \citealt{1989PASJ...41....1N}). However, the long-term spin derivative is only a qualitative estimate because there are many short-term changes in this supergiant accretion system. To obtain the precise spin history of this source, future high cadence observations are required.

\begin{figure}
\centering
\includegraphics[width=0.5\textwidth]{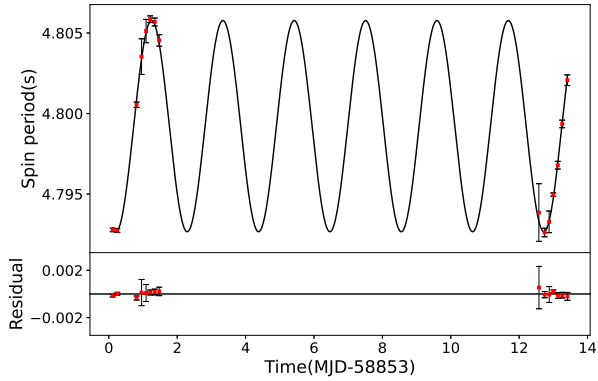}
\caption{The observed spin period values of Cen X-3 during the epoch from MJD 58853 to MJD 58865 based on Insight-HXMT in 2020. The black solid curve represents the best fitting model due to the Doppler motion of the binary. Residuals are shown in the lower panel.}
\label{fig:orbit00}
\end{figure}

Combing the two observations in 2020 together, we couldn't find a more precise value for all the free parameters with the fitting plots shown in Figure\ \ref{fig:orbit00}, and the fitted values are presented in Table \ref{tab:orbital}. Therefore, we use the measurement on January 05, 2020 due to smaller error bars in the following analysis. As summarised in Table \ref{tab:orbital}, we obtain an accurate intrinsic spin period of $4.79920 \pm 0.00006$ s, an eclipse epoch $T_{\rm ecl}$ of 58852.697 $\pm$ 0.006 d, and the projected semi-major axis of about 39.76 $\pm$ 0.41 light-sec at MJD 58852 which is a little larger than earlier measurements of 39.6612 $\pm$ 0.0009 light-sec by \cite{2010MNRAS.401.1532R}, but consistent within error bars.

\begin{table*}
    \centering
    \caption{Measured values of the spin period and orbital parameters of Cen X-3.}
    \label{tab:orbital}
    \renewcommand\arraystretch{1.5}
    \setlength{\tabcolsep}{0.18mm}{
    \begin{tabular}{l c c c c}
    \hline \hline 
    \multirow{2}{*}{Parameters} & \multirow{2}{*}{February 04, 2018} & \multirow{2}{*}{January 05, 2020} & \multirow{2}{*}{January 17, 2020} & \multirow{2}{*}{January 05/17, 2020} \\ \\
    \hline 
    $p_{0}$ (s) & $4.80049 \pm 0.00010$ & $4.79920 \pm 0.00006$ &  $4.79875 \pm 0.00031$   &  $4.79922\pm 0.00007$ \\ 
    $a_{x}\sin i$ (lt-s) & 36.65$\pm$ 1.18 & 39.76$\pm$ 0.41  & 36.36 $\pm$ 1.15 &   39.22 $\pm$ 0.47\\ 
    $T_{\rm ecl}$ (d) & 58153.537  $\pm$0.009 &  58852.697 $\pm$0.006 &  58865.183 $\pm$0.017 &  58852.685 $\pm$0.006 \\  \hline
    $^a P_{\rm orb}$ (d) & - & 2.08695634 $\pm$ 0.00000001 &- &-\\
    $\dot{P}_{\rm orb}/P_{\rm orb}$ (yr$^{-1}$) & - & -(1.7832 $\pm$ 0.0001) $\times$ 10$^{-6}$ &-&-\\ 
    \hline 
    \end{tabular} 
    \begin{tablenotes}
    \item $^a$ orbital period referred to the epoch of MJD 58852.697.
    \end{tablenotes}
    }
\end{table*}

\begin{figure}
    \centering
    \includegraphics[width=.5\textwidth]{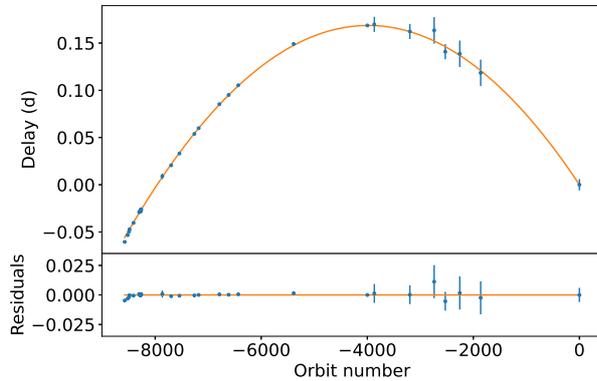}
    \caption{The observed mid-eclipse times, T$_{\rm eclipse}$, are shown as a function of the orbit number. The latest data point is from this work. The solid curve has a quadratic trend and represents the best-fit model for orbital decay. The lower panel shows the residuals from the best fit.}
    \label{fig:orbital_fit}
\end{figure}

In order to determine the accurate orbital period and its derivative, we combined the new measurement of the epoch of eclipse made by extensive observations of HXMT on January 05, 2020 with previously obtained epochs, as seen in Table\ \ref{tab:derivative}. The $n$th mid-eclipse epochs and the orbital change satisfy the quadratic function:
\begin{equation}
T_{n}^{\rm ecl}=T_{0}^{\rm ecl}+n P_{\mathrm{orb}}+\frac{1}{2} n^{2} P_{\mathrm{orb}} \dot{P}_{\mathrm{orb}}
\end{equation}
Here $P_{\mathrm{orb}}$ and $\dot{P}_{\mathrm{orb}}$ are the orbital period and the period derivative at the epoch $T_{0}^{\rm ecl}$. The best-fit plots including the residuals of mid-eclipse time from fitting are shown in Figure\ \ref{fig:orbital_fit}. These fits are performed by a least square minimization and are acceptable. The new orbital period and its decay are reported in Table\ \ref{tab:orbital}. We obtain the orbital period of 2.08695634 $\pm$ 0.00000001 d and orbital decay $\dot{P}_{\rm orb}/P_{\rm orb}$ of -(1.7832 $\pm$ 0.0001) $\times$ 10$^{-6}$ yr$^{-1}$ referred to the epoch of MJD 58852.697, which is consistent with the one obtained by \cite{2015AA...577A.130F} but more precise by a factor of 10. 

\begin{table}
\begin{centering}
\footnotesize
\caption{The orbital epoch history in Cen X-3.}
\label{tab:derivative}
\begin{tabular}{lll}
\hline
\hline
\multirow{2}{*}{Epoch Time} &  \multirow{2}{*}{Satellite} & \multirow{2}{*}{Reference}\\ 
\multirow{2}{*}{T$_{\rm eclipse}$ MJD}  &  & \\
\\
\hline
\hline
40958.34643(45) & Uhuru & \citet{1977ApJ...214..235F}\\
41077.31497(15) &  Uhuru & \citet{1977ApJ...214..235F}\\
41131.58181(29) &  Uhuru & \citet{1977ApJ...214..235F}\\
41148.28051(16) & Uhuru & \citet{1977ApJ...214..235F}\\
41304.81533(14) &  Uhuru & \citet{1977ApJ...214..235F}\\
41528.1401(3) &  Uhuru & \citet{1977ApJ...214..235F}\\
41551.09798(17) &  Uhuru & \citet{1977ApJ...214..235F}\\
41569.88199(11) &  Uhuru & \citet{1977ApJ...214..235F}\\
41574.05610(13) &  Uhuru & \citet{1977ApJ...214..235F}\\
41576.1433(1) &  Uhuru & \citet{1977ApJ...214..235F}\\
41578.23037(7) &   Uhuru & \citet{1977ApJ...214..235F}\\
41580.31722(9) &  Uhuru & \citet{1977ApJ...214..235F}\\
41584.49193(10) &  Uhuru & \citet{1977ApJ...214..235F}\\
41590.75328(15) &  Uhuru & \citet{1977ApJ...214..235F}\\
41592.84025(15) &  Uhuru & \citet{1977ApJ...214..235F}\\
41599.10212(15) &  Uhuru & \citet{1977ApJ...214..235F}\\
41601.18930(14) &  Uhuru & \citet{1977ApJ...214..235F}\\
41603.27671(21) &  Uhuru & \citet{1977ApJ...214..235F}\\
42438.128(3) & {Ariel-V} & \citet{1976MNRAS.174P..45T} \\
42786.6755(7)  & {Cos-B} & \citet{1980A} \\
43112.26642(40)  & {SAS-3} & \citet{1983ApJ...264..568K} \\
43700.83275(43)  & {HEAO-1} & \citet{1983ApJ...272..678H}\\
43869.88910(2)  & {SAS-3} & \citet{1983ApJ...264..568K}\\
44685.94760(5)  & {Hakucho} & \citet{1983ApJ...264..563M}\\
45049.1025(1)  & {Hakucho} & \citet{1984PASJ...36..667N}\\
45428.95421(5)  & {Tenma} & \citet{1984PASJ...36..667N}\\
47607.8688(8)  & {Ginga} & \citet{1992ApJ...396..147N}\\
50506.788423(7)  & {RXTE} & \citet{2010MNRAS.401.1532R}\\
50782.279(8) & RXTE/ASM & \citet{2015AA...577A.130F} \\
52180.589(8) & RXTE/ASM & \citet{2015AA...577A.130F} \\
53136.455(14)  & {INTEGRAL} & \citet{2015AA...577A.130F} \\
53574.711(8) & RXTE/ASM & \citet{2015AA...577A.130F} \\
54144.471(14)  & {INTEGRAL} & \citet{2015AA...577A.130F}\\
54966.745(14) & RXTE/ASM & \citet{2015AA...577A.130F} \\
58852.697(6) & Insight-HXMT & present work \\

\hline
\end{tabular}
\end{centering}
\end{table}

\section{Conclusion} \label{sec:summary}

We have carried out a detailed temporal analysis for Cen X-3 with observations by HXMT in 2018 and 2020. The energy-dependent pulse profiles show a double-peaked feature at the energies below $\sim$ 10 keV with an apparent second peak, and a single main peak in higher energies. In addition, pulse profiles at low energies vary over time in 2018, while stably show a single peak in 2020, which may suggest the different emission patterns between 2018 and 2020, i.e., the change of the emission pattern from the mixture of a pencil and a fan beam to a pencil-like beam. The intrinsic spin period after binary orbit correction of the X-ray pulsar is found to be $4.79920 \pm 0.00006$ s at MJD 58852.697, with a spin-up trend from 2018 -- 2020. The orbital parameters of Cen X-3 are fitted with the pulse period modulation. We obtain an updated orbital period and an improved orbital decay rate. The orbital period and its decay rate are determined to be 2.08695634 $\pm$ 0.00000001 d and -(1.7832 $\pm$ 0.0001) $\times$ 10$^{-6}$ yr$^{-1}$ with high precision, respectively, as referred to the epoch of MJD 58852.697.

\section*{Acknowledgements}
We are grateful to the referee for the useful comments to improve the manuscript. This work is supported by the National Key Research and Development Program of China (Grants No. 2021YFA0718503), the NSFC (12133007, U1838103). This work has made use of data from the \textit{Insight}-HXMT mission, a project funded by China National Space Administration (CNSA) and the Chinese Academy of Sciences (CAS).

\bibliographystyle{harv}

\bibliography{refer}

\begin{thebibliography}{41}
\expandafter\ifx\csname natexlab\endcsname\relax\def\natexlab#1{#1}\fi
\providecommand{\url}[1]{\texttt{#1}}
\providecommand{\href}[2]{#2}
\providecommand{\path}[1]{#1}
\providecommand{\DOIprefix}{doi:}
\providecommand{\ArXivprefix}{arXiv:}
\providecommand{\URLprefix}{URL: }
\providecommand{\Pubmedprefix}{pmid:}
\providecommand{\doi}[1]{\href{http://dx.doi.org/#1}{\path{#1}}}
\providecommand{\Pubmed}[1]{\href{pmid:#1}{\path{#1}}}
\providecommand{\bibinfo}[2]{#2}
\ifx\xfnm\relax \def\xfnm[#1]{\unskip,\space#1}\fi
\bibitem[{{Arnaud}(1996)}]{1996ASPC..101...17A}
\bibinfo{author}{{Arnaud}, K.A.}, \bibinfo{year}{1996}.
\newblock \bibinfo{title}{{XSPEC: The First Ten Years}}, in:
  \bibinfo{editor}{{Jacoby}, G.H.}, \bibinfo{editor}{{Barnes}, J.} (Eds.),
  \bibinfo{booktitle}{Astronomical Data Analysis Software and Systems V},
  p.~\bibinfo{pages}{17}.
\bibitem[{{Ash} et~al.(1999){Ash}, {Reynolds}, {Roche}, {Norton}, {Still} and
  {Morales-Rueda}}]{1999MNRAS.307..357A}
\bibinfo{author}{{Ash}, T.D.C.}, \bibinfo{author}{{Reynolds}, A.P.},
  \bibinfo{author}{{Roche}, P.}, \bibinfo{author}{{Norton}, A.J.},
  \bibinfo{author}{{Still}, M.D.}, \bibinfo{author}{{Morales-Rueda}, L.},
  \bibinfo{year}{1999}.
\newblock \bibinfo{title}{{The mass of the neutron star in Centaurus X-3}}.
\newblock \bibinfo{journal}{\mnras} \bibinfo{volume}{307},
  \bibinfo{pages}{357--364}.
\newblock \DOIprefix\doi{10.1046/j.1365-8711.1999.02605.x}.
\bibitem[{{Basko} and {Sunyaev}(1975)}]{1975A&A....42..311B}
\bibinfo{author}{{Basko}, M.M.}, \bibinfo{author}{{Sunyaev}, R.A.},
  \bibinfo{year}{1975}.
\newblock \bibinfo{title}{{Radiative transfer in a strong magnetic field and
  accreting X-ray pulsars.}}
\newblock \bibinfo{journal}{\aap} \bibinfo{volume}{42},
  \bibinfo{pages}{311--321}.
\bibitem[{{Basko} and {Sunyaev}(1976)}]{1976MNRAS.175..395B}
\bibinfo{author}{{Basko}, M.M.}, \bibinfo{author}{{Sunyaev}, R.A.},
  \bibinfo{year}{1976}.
\newblock \bibinfo{title}{{The limiting luminosity of accreting neutron stars
  with magnetic fields.}}
\newblock \bibinfo{journal}{\mnras} \bibinfo{volume}{175},
  \bibinfo{pages}{395--417}.
\newblock \DOIprefix\doi{10.1093/mnras/175.2.395}.
\bibitem[{{Becker} and {Wolff}(2007)}]{2007ApJ...654..435B}
\bibinfo{author}{{Becker}, P.A.}, \bibinfo{author}{{Wolff}, M.T.},
  \bibinfo{year}{2007}.
\newblock \bibinfo{title}{{Thermal and Bulk Comptonization in Accretion-powered
  X-Ray Pulsars}}.
\newblock \bibinfo{journal}{\apj} \bibinfo{volume}{654},
  \bibinfo{pages}{435--457}.
\newblock \DOIprefix\doi{10.1086/509108},
  \href{http://arxiv.org/abs/astro-ph/0609035}{{\tt arXiv:astro-ph/0609035}}.
\bibitem[{{Burderi} et~al.(2000){Burderi}, {Di Salvo}, {Robba}, {La Barbera}
  and {Guainazzi}}]{2000ApJ...530..429B}
\bibinfo{author}{{Burderi}, L.}, \bibinfo{author}{{Di Salvo}, T.},
  \bibinfo{author}{{Robba}, N.R.}, \bibinfo{author}{{La Barbera}, A.},
  \bibinfo{author}{{Guainazzi}, M.}, \bibinfo{year}{2000}.
\newblock \bibinfo{title}{{The 0.1-100 KEV Spectrum of Centaurus X-3: Pulse
  Phase Spectroscopy of the Cyclotron Line and Magnetic Field Structure}}.
\newblock \bibinfo{journal}{\apj} \bibinfo{volume}{530},
  \bibinfo{pages}{429--440}.
\newblock \DOIprefix\doi{10.1086/308336}.
\bibitem[{{Chen} et~al.(2022){Chen}, {Wang}, {You}, {Tian}, {Liu}, {Zhang},
  {Ding}, {Qu}, {Zhang}, {Song}, {Lu} and {Zhang}}]{2022MNRAS.513.4875C}
\bibinfo{author}{{Chen}, X.}, \bibinfo{author}{{Wang}, W.},
  \bibinfo{author}{{You}, B.}, \bibinfo{author}{{Tian}, P.F.},
  \bibinfo{author}{{Liu}, Q.}, \bibinfo{author}{{Zhang}, P.},
  \bibinfo{author}{{Ding}, Y.Z.}, \bibinfo{author}{{Qu}, J.L.},
  \bibinfo{author}{{Zhang}, S.N.}, \bibinfo{author}{{Song}, L.M.},
  \bibinfo{author}{{Lu}, F.J.}, \bibinfo{author}{{Zhang}, S.},
  \bibinfo{year}{2022}.
\newblock \bibinfo{title}{{Wavelet analysis of MAXI J1535-571 with
  Insight-HXMT}}.
\newblock \bibinfo{journal}{\mnras} \bibinfo{volume}{513},
  \bibinfo{pages}{4875--4886}.
\newblock \DOIprefix\doi{10.1093/mnras/stac1175},
  \href{http://arxiv.org/abs/2204.12030}{{\tt arXiv:2204.12030}}.
\bibitem[{{Ding} et~al.(2021){Ding}, {Wang}, {Zhang}, {Bu}, {Cai}, {Cao},
  {Zhi}, {Chen}, {Chen}, {Chen}, {Chen}, {Chen}, {Cui}, {Du}, {Gao}, {Gao},
  {Ge}, {Gu}, {Guan}, {Guo}, {Han}, {Huang}, {Huo}, {Jia}, {Jiang}, {Jin},
  {Kong}, {Li}, {Li}, {Li}, {Li}, {Li}, {Li}, {Li}, {Li}, {Li}, {Liang},
  {Liao}, {Liu}, {Liu}, {Liu}, {Liu}, {Liu}, {Lu}, {Lu}, {Lou}, {Luo}, {Ma},
  {Ma}, {Meng}, {Nang}, {Nie}, {Qu}, {Ren}, {Sai}, {Song}, {Song}, {Sun},
  {Tan}, {Tao}, {Tuo}, {Wang}, {Wang}, {Wang}, {Wang}, {Wang}, {Wen}, {Wu},
  {Wu}, {Wu}, {Xiao}, {Xiao}, {Xiong}, {Xu}, {Yang}, {Yang}, {Yang}, {Yi},
  {Yin}, {You}, {Zhang}, {Zhang}, {Zhang}, {Zhang}, {Zhang}, {Zhang}, {Zhang},
  {Zhang}, {Zhang}, {Zhang}, {Zhao}, {Zhao}, {Zheng}, {Zheng} and
  {Zhou}}]{2021MNRAS.503.6045D}
\bibinfo{author}{{Ding}, Y.Z.}, \bibinfo{author}{{Wang}, W.},
  \bibinfo{author}{{Zhang}, P.}, \bibinfo{author}{{Bu}, Q.C.},
  \bibinfo{author}{{Cai}, C.}, \bibinfo{author}{{Cao}, X.L.},
  \bibinfo{author}{{Zhi}, C.}, \bibinfo{author}{{Chen}, L.},
  \bibinfo{author}{{Chen}, T.X.}, \bibinfo{author}{{Chen}, Y.B.},
  \bibinfo{author}{{Chen}, Y.}, \bibinfo{author}{{Chen}, Y.P.},
  \bibinfo{author}{{Cui}, W.W.}, \bibinfo{author}{{Du}, Y.Y.},
  \bibinfo{author}{{Gao}, G.H.}, \bibinfo{author}{{Gao}, H.},
  \bibinfo{author}{{Ge}, M.Y.}, \bibinfo{author}{{Gu}, Y.D.},
  \bibinfo{author}{{Guan}, J.}, \bibinfo{author}{{Guo}, C.C.},
  \bibinfo{author}{{Han}, D.W.}, \bibinfo{author}{{Huang}, Y.},
  \bibinfo{author}{{Huo}, J.}, \bibinfo{author}{{Jia}, S.M.},
  \bibinfo{author}{{Jiang}, W.C.}, \bibinfo{author}{{Jin}, J.},
  \bibinfo{author}{{Kong}, L.D.}, \bibinfo{author}{{Li}, B.},
  \bibinfo{author}{{Li}, C.K.}, \bibinfo{author}{{Li}, G.},
  \bibinfo{author}{{Li}, T.P.}, \bibinfo{author}{{Li}, W.},
  \bibinfo{author}{{Li}, X.}, \bibinfo{author}{{Li}, X.B.},
  \bibinfo{author}{{Li}, X.F.}, \bibinfo{author}{{Li}, Z.W.},
  \bibinfo{author}{{Liang}, X.H.}, \bibinfo{author}{{Liao}, J.Y.},
  \bibinfo{author}{{Liu}, B.S.}, \bibinfo{author}{{Liu}, C.Z.},
  \bibinfo{author}{{Liu}, H.X.}, \bibinfo{author}{{Liu}, H.W.},
  \bibinfo{author}{{Liu}, X.J.}, \bibinfo{author}{{Lu}, F.J.},
  \bibinfo{author}{{Lu}, X.F.}, \bibinfo{author}{{Lou}, Q.},
  \bibinfo{author}{{Luo}, T.}, \bibinfo{author}{{Ma}, R.C.},
  \bibinfo{author}{{Ma}, X.}, \bibinfo{author}{{Meng}, B.},
  \bibinfo{author}{{Nang}, Y.}, \bibinfo{author}{{Nie}, J.Y.},
  \bibinfo{author}{{Qu}, J.L.}, \bibinfo{author}{{Ren}, X.Q.},
  \bibinfo{author}{{Sai}, N.}, \bibinfo{author}{{Song}, L.M.},
  \bibinfo{author}{{Song}, X.Y.}, \bibinfo{author}{{Sun}, L.},
  \bibinfo{author}{{Tan}, Y.}, \bibinfo{author}{{Tao}, L.},
  \bibinfo{author}{{Tuo}, Y.L.}, \bibinfo{author}{{Wang}, C.},
  \bibinfo{author}{{Wang}, L.J.}, \bibinfo{author}{{Wang}, P.J.},
  \bibinfo{author}{{Wang}, W.S.}, \bibinfo{author}{{Wang}, Y.S.},
  \bibinfo{author}{{Wen}, X.Y.}, \bibinfo{author}{{Wu}, B.Y.},
  \bibinfo{author}{{Wu}, B.B.}, \bibinfo{author}{{Wu}, M.},
  \bibinfo{author}{{Xiao}, G.C.}, \bibinfo{author}{{Xiao}, S.},
  \bibinfo{author}{{Xiong}, S.L.}, \bibinfo{author}{{Xu}, Y.P.},
  \bibinfo{author}{{Yang}, R.J.}, \bibinfo{author}{{Yang}, S.},
  \bibinfo{author}{{Yang}, Y.J.}, \bibinfo{author}{{Yi}, Q.B.},
  \bibinfo{author}{{Yin}, Q.Q.}, \bibinfo{author}{{You}, Y.},
  \bibinfo{author}{{Zhang}, F.}, \bibinfo{author}{{Zhang}, H.M.},
  \bibinfo{author}{{Zhang}, J.}, \bibinfo{author}{{Zhang}, P.},
  \bibinfo{author}{{Zhang}, S.}, \bibinfo{author}{{Zhang}, S.N.},
  \bibinfo{author}{{Zhang}, W.C.}, \bibinfo{author}{{Zhang}, W.},
  \bibinfo{author}{{Zhang}, Y.F.}, \bibinfo{author}{{Zhang}, Y.H.},
  \bibinfo{author}{{Zhao}, H.S.}, \bibinfo{author}{{Zhao}, X.F.},
  \bibinfo{author}{{Zheng}, S.J.}, \bibinfo{author}{{Zheng}, Y.G.},
  \bibinfo{author}{{Zhou}, D.K.}, \bibinfo{year}{2021}.
\newblock \bibinfo{title}{{QPOs and orbital elements of X-ray binary 4U 0115+63
  during the 2017 outburst observed by Insight-HXMT}}.
\newblock \bibinfo{journal}{\mnras} \bibinfo{volume}{503},
  \bibinfo{pages}{6045--6058}.
\newblock \DOIprefix\doi{10.1093/mnras/stab835},
  \href{http://arxiv.org/abs/2102.09498}{{\tt arXiv:2102.09498}}.
\bibitem[{{Fabbiano} and {Schreier}(1977)}]{1977ApJ...214..235F}
\bibinfo{author}{{Fabbiano}, G.}, \bibinfo{author}{{Schreier}, E.J.},
  \bibinfo{year}{1977}.
\newblock \bibinfo{title}{{Further studies of the pulsation period and orbital
  elements of Centaurus X-3.}}
\newblock \bibinfo{journal}{\apj} \bibinfo{volume}{214},
  \bibinfo{pages}{235--244}.
\newblock \DOIprefix\doi{10.1086/155247}.
\bibitem[{{Falanga} et~al.(2015){Falanga}, {Bozzo}, {Lutovinov},
  {Bonnet-Bidaud}, {Fetisova} and {Puls}}]{2015AA...577A.130F}
\bibinfo{author}{{Falanga}, M.}, \bibinfo{author}{{Bozzo}, E.},
  \bibinfo{author}{{Lutovinov}, A.}, \bibinfo{author}{{Bonnet-Bidaud}, J.M.},
  \bibinfo{author}{{Fetisova}, Y.}, \bibinfo{author}{{Puls}, J.},
  \bibinfo{year}{2015}.
\newblock \bibinfo{title}{{Ephemeris, orbital decay, and masses of ten
  eclipsing high-mass X-ray binaries}}.
\newblock \bibinfo{journal}{\aap} \bibinfo{volume}{577}, \bibinfo{pages}{A130}.
\newblock \DOIprefix\doi{10.1051/0004-6361/201425191},
  \href{http://arxiv.org/abs/1502.07126}{{\tt arXiv:1502.07126}}.
\bibitem[{{Giacconi} et~al.(1971){Giacconi}, {Gursky}, {Kellogg}, {Schreier}
  and {Tananbaum}}]{1971ApJ...167L..67G}
\bibinfo{author}{{Giacconi}, R.}, \bibinfo{author}{{Gursky}, H.},
  \bibinfo{author}{{Kellogg}, E.}, \bibinfo{author}{{Schreier}, E.},
  \bibinfo{author}{{Tananbaum}, H.}, \bibinfo{year}{1971}.
\newblock \bibinfo{title}{{Discovery of Periodic X-Ray Pulsations in Centaurus
  X-3 from UHURU}}.
\newblock \bibinfo{journal}{\apjl} \bibinfo{volume}{167}, \bibinfo{pages}{L67}.
\newblock \DOIprefix\doi{10.1086/180762}.
\bibitem[{{Howe} et~al.(1983){Howe}, {Primini}, {Bautz}, {Lang}, {Levine} and
  {Lewin}}]{1983ApJ...272..678H}
\bibinfo{author}{{Howe}, S.K.}, \bibinfo{author}{{Primini}, F.A.},
  \bibinfo{author}{{Bautz}, M.W.}, \bibinfo{author}{{Lang}, F.L.},
  \bibinfo{author}{{Levine}, A.M.}, \bibinfo{author}{{Lewin}, W.H.G.},
  \bibinfo{year}{1983}.
\newblock \bibinfo{title}{{HEAO 1 high-energy X-ray observations of Centaurus
  X-3.}}
\newblock \bibinfo{journal}{\apj} \bibinfo{volume}{272},
  \bibinfo{pages}{678--686}.
\newblock \DOIprefix\doi{10.1086/161330}.
\bibitem[{{Huppenkothen} et~al.(2019){Huppenkothen}, {Bachetti}, {Stevens},
  {Migliari}, {Balm}, {Hammad}, {Khan}, {Mishra}, {Rashid}, {Sharma}, {Martinez
  Ribeiro} and {Valles Blanco}}]{2019ApJ...881...39H}
\bibinfo{author}{{Huppenkothen}, D.}, \bibinfo{author}{{Bachetti}, M.},
  \bibinfo{author}{{Stevens}, A.L.}, \bibinfo{author}{{Migliari}, S.},
  \bibinfo{author}{{Balm}, P.}, \bibinfo{author}{{Hammad}, O.},
  \bibinfo{author}{{Khan}, U.M.}, \bibinfo{author}{{Mishra}, H.},
  \bibinfo{author}{{Rashid}, H.}, \bibinfo{author}{{Sharma}, S.},
  \bibinfo{author}{{Martinez Ribeiro}, E.}, \bibinfo{author}{{Valles Blanco},
  R.}, \bibinfo{year}{2019}.
\newblock \bibinfo{title}{{Stingray: A Modern Python Library for Spectral
  Timing}}.
\newblock \bibinfo{journal}{\apj} \bibinfo{volume}{881}, \bibinfo{pages}{39}.
\newblock \DOIprefix\doi{10.3847/1538-4357/ab258d},
  \href{http://arxiv.org/abs/1901.07681}{{\tt arXiv:1901.07681}}.
\bibitem[{{Hutchings} et~al.(1979){Hutchings}, {Cowley}, {Crampton}, {van
  Paradijs} and {White}}]{1979ApJ...229.1079H}
\bibinfo{author}{{Hutchings}, J.B.}, \bibinfo{author}{{Cowley}, A.P.},
  \bibinfo{author}{{Crampton}, D.}, \bibinfo{author}{{van Paradijs}, J.},
  \bibinfo{author}{{White}, N.E.}, \bibinfo{year}{1979}.
\newblock \bibinfo{title}{{Centaurus X-3.}}
\newblock \bibinfo{journal}{\apj} \bibinfo{volume}{229},
  \bibinfo{pages}{1079--1084}.
\newblock \DOIprefix\doi{10.1086/157042}.
\bibitem[{{Kelley} et~al.(1983){Kelley}, {Jernigan}, {Levine}, {Petro} and
  {Rappaport}}]{1983ApJ...264..568K}
\bibinfo{author}{{Kelley}, R.L.}, \bibinfo{author}{{Jernigan}, J.G.},
  \bibinfo{author}{{Levine}, A.}, \bibinfo{author}{{Petro}, L.D.},
  \bibinfo{author}{{Rappaport}, S.}, \bibinfo{year}{1983}.
\newblock \bibinfo{title}{{Discovery of 13.5 S X-ray pulsations from LMC X-4
  and an orbital determination.}}
\newblock \bibinfo{journal}{\apj} \bibinfo{volume}{264},
  \bibinfo{pages}{568--574}.
\newblock \DOIprefix\doi{10.1086/160626}.
\bibitem[{{Krzeminski}(1974)}]{1974ApJ...192L.135K}
\bibinfo{author}{{Krzeminski}, W.}, \bibinfo{year}{1974}.
\newblock \bibinfo{title}{{The Identification and UBV Photometry of the Visible
  Component of the Centaurus X-3 Binary System}}.
\newblock \bibinfo{journal}{\apjl} \bibinfo{volume}{192},
  \bibinfo{pages}{L135}.
\newblock \DOIprefix\doi{10.1086/181609}.
\bibitem[{{Leahy} et~al.(1983){Leahy}, {Elsner} and
  {Weisskopf}}]{1983ApJ...272..256L}
\bibinfo{author}{{Leahy}, D.A.}, \bibinfo{author}{{Elsner}, R.F.},
  \bibinfo{author}{{Weisskopf}, M.C.}, \bibinfo{year}{1983}.
\newblock \bibinfo{title}{{On searches for periodic pulsed emission - The
  Rayleigh test compared to epoch folding}}.
\newblock \bibinfo{journal}{\apj} \bibinfo{volume}{272},
  \bibinfo{pages}{256--258}.
\newblock \DOIprefix\doi{10.1086/161288}.
\bibitem[{{Li} et~al.(2012){Li}, {Wang} and {Zhao}}]{2012MNRAS.423.2854L}
\bibinfo{author}{{Li}, J.}, \bibinfo{author}{{Wang}, W.},
  \bibinfo{author}{{Zhao}, Y.}, \bibinfo{year}{2012}.
\newblock \bibinfo{title}{{Cyclotron resonance energies and orbital elements of
  accretion pulsar 4U 0115+63 during the giant outburst in 2008}}.
\newblock \bibinfo{journal}{\mnras} \bibinfo{volume}{423},
  \bibinfo{pages}{2854--2867}.
\newblock \DOIprefix\doi{10.1111/j.1365-2966.2012.21096.x},
  \href{http://arxiv.org/abs/1204.2908}{{\tt arXiv:1204.2908}}.
\bibitem[{{Liu} et~al.(2022){Liu}, {Wang}, {Chen}, {Yang}, {Lu}, {Song}, {Qu},
  {Zhang} and {Zhang}}]{2022MNRAS.516.5579L}
\bibinfo{author}{{Liu}, Q.}, \bibinfo{author}{{Wang}, W.},
  \bibinfo{author}{{Chen}, X.}, \bibinfo{author}{{Yang}, W.},
  \bibinfo{author}{{Lu}, F.J.}, \bibinfo{author}{{Song}, L.M.},
  \bibinfo{author}{{Qu}, J.L.}, \bibinfo{author}{{Zhang}, S.},
  \bibinfo{author}{{Zhang}, S.N.}, \bibinfo{year}{2022}.
\newblock \bibinfo{title}{{Detection of a quasi-periodic oscillation at 40 mHz
  in Cen X-3 with Insight-HXMT}}.
\newblock \bibinfo{journal}{\mnras} \bibinfo{volume}{516},
  \bibinfo{pages}{5579--5587}.
\newblock \DOIprefix\doi{10.1093/mnras/stac2646},
  \href{http://arxiv.org/abs/2209.06662}{{\tt arXiv:2209.06662}}.
\bibitem[{{Murakami} et~al.(1983){Murakami}, {Inoue}, {Kawai}, {Koyama},
  {Makishima}, {Matsuoka}, {Oda}, {Ogawara}, {Ohashi}, {Shibazaki}, {Tanaka},
  {Hayakawa}, {Kunieda}, {Makino}, {Masai}, {Nagase}, {Tawara}, {Miyamoto},
  {Tsunemi}, {Yamashita} and {Kondo}}]{1983ApJ...264..563M}
\bibinfo{author}{{Murakami}, T.}, \bibinfo{author}{{Inoue}, H.},
  \bibinfo{author}{{Kawai}, N.}, \bibinfo{author}{{Koyama}, K.},
  \bibinfo{author}{{Makishima}, K.}, \bibinfo{author}{{Matsuoka}, M.},
  \bibinfo{author}{{Oda}, M.}, \bibinfo{author}{{Ogawara}, Y.},
  \bibinfo{author}{{Ohashi}, T.}, \bibinfo{author}{{Shibazaki}, N.},
  \bibinfo{author}{{Tanaka}, Y.}, \bibinfo{author}{{Hayakawa}, S.},
  \bibinfo{author}{{Kunieda}, H.}, \bibinfo{author}{{Makino}, F.},
  \bibinfo{author}{{Masai}, K.}, \bibinfo{author}{{Nagase}, F.},
  \bibinfo{author}{{Tawara}, Y.}, \bibinfo{author}{{Miyamoto}, S.},
  \bibinfo{author}{{Tsunemi}, H.}, \bibinfo{author}{{Yamashita}, K.},
  \bibinfo{author}{{Kondo}, I.}, \bibinfo{year}{1983}.
\newblock \bibinfo{title}{{Observation of Centaurus X-3 by HAKUCHO.}}
\newblock \bibinfo{journal}{\apj} \bibinfo{volume}{264},
  \bibinfo{pages}{563--567}.
\newblock \DOIprefix\doi{10.1086/160625}.
\bibitem[{{Nagase}(1989)}]{1989PASJ...41....1N}
\bibinfo{author}{{Nagase}, F.}, \bibinfo{year}{1989}.
\newblock \bibinfo{title}{{Accretion-powered X-ray pulsars.}}
\newblock \bibinfo{journal}{\pasj} \bibinfo{volume}{41}, \bibinfo{pages}{1}.
\bibitem[{{Nagase} et~al.(1992){Nagase}, {Corbet}, {Day}, {Inoue}, {Takeshima},
  {Yoshida} and {Mihara}}]{1992ApJ...396..147N}
\bibinfo{author}{{Nagase}, F.}, \bibinfo{author}{{Corbet}, R.H.D.},
  \bibinfo{author}{{Day}, C.S.R.}, \bibinfo{author}{{Inoue}, H.},
  \bibinfo{author}{{Takeshima}, T.}, \bibinfo{author}{{Yoshida}, K.},
  \bibinfo{author}{{Mihara}, T.}, \bibinfo{year}{1992}.
\newblock \bibinfo{title}{{GINGA Observations of Centaurus X-3}}.
\newblock \bibinfo{journal}{\apj} \bibinfo{volume}{396}, \bibinfo{pages}{147}.
\newblock \DOIprefix\doi{10.1086/171705}.
\bibitem[{{Nagase} et~al.(1984){Nagase}, {Hayakawa}, {Kii}, {Sato}, {Ikegami},
  {Kawai}, {Makishima}, {Matsuoka}, {Mitani}, {Murakami}, {Oda}, {Ohashi},
  {Tanaka} and {Kitamoto}}]{1984PASJ...36..667N}
\bibinfo{author}{{Nagase}, F.}, \bibinfo{author}{{Hayakawa}, S.},
  \bibinfo{author}{{Kii}, T.}, \bibinfo{author}{{Sato}, N.},
  \bibinfo{author}{{Ikegami}, T.}, \bibinfo{author}{{Kawai}, N.},
  \bibinfo{author}{{Makishima}, K.}, \bibinfo{author}{{Matsuoka}, M.},
  \bibinfo{author}{{Mitani}, K.}, \bibinfo{author}{{Murakami}, T.},
  \bibinfo{author}{{Oda}, M.}, \bibinfo{author}{{Ohashi}, T.},
  \bibinfo{author}{{Tanaka}, Y.}, \bibinfo{author}{{Kitamoto}, S.},
  \bibinfo{year}{1984}.
\newblock \bibinfo{title}{{Pulse-period changes of X-ray pulsars measured with
  Hakucho and Tenma.}}
\newblock \bibinfo{journal}{\pasj} \bibinfo{volume}{36},
  \bibinfo{pages}{667--678}.
\bibitem[{{Poutanen} et~al.(2013){Poutanen}, {Mushtukov}, {Suleimanov},
  {Tsygankov}, {Nagirner}, {Doroshenko} and {Lutovinov}}]{2013ApJ...777..115P}
\bibinfo{author}{{Poutanen}, J.}, \bibinfo{author}{{Mushtukov}, A.A.},
  \bibinfo{author}{{Suleimanov}, V.F.}, \bibinfo{author}{{Tsygankov}, S.S.},
  \bibinfo{author}{{Nagirner}, D.I.}, \bibinfo{author}{{Doroshenko}, V.},
  \bibinfo{author}{{Lutovinov}, A.A.}, \bibinfo{year}{2013}.
\newblock \bibinfo{title}{{A Reflection Model for the Cyclotron Lines in the
  Spectra of X-Ray Pulsars}}.
\newblock \bibinfo{journal}{\apj} \bibinfo{volume}{777}, \bibinfo{pages}{115}.
\newblock \DOIprefix\doi{10.1088/0004-637X/777/2/115},
  \href{http://arxiv.org/abs/1304.2633}{{\tt arXiv:1304.2633}}.
\bibitem[{{Raichur} and {Paul}(2008a)}]{2008MNRAS.387..439R}
\bibinfo{author}{{Raichur}, H.}, \bibinfo{author}{{Paul}, B.},
  \bibinfo{year}{2008}a.
\newblock \bibinfo{title}{{Long-term flux variations in Cen X-3: clues from
  flux-dependent orbital modulation and pulsed fraction}}.
\newblock \bibinfo{journal}{\mnras} \bibinfo{volume}{387},
  \bibinfo{pages}{439--445}.
\newblock \DOIprefix\doi{10.1111/j.1365-2966.2008.13251.x},
  \href{http://arxiv.org/abs/0804.1614}{{\tt arXiv:0804.1614}}.
\bibitem[{{Raichur} and {Paul}(2008b)}]{2008ApJ...685.1109R}
\bibinfo{author}{{Raichur}, H.}, \bibinfo{author}{{Paul}, B.},
  \bibinfo{year}{2008}b.
\newblock \bibinfo{title}{{Quasi-Periodic Oscillations in Cen X-3 and the
  Long-Term Intensity Variations}}.
\newblock \bibinfo{journal}{\apj} \bibinfo{volume}{685},
  \bibinfo{pages}{1109--1113}.
\newblock \DOIprefix\doi{10.1086/591037},
  \href{http://arxiv.org/abs/0806.0949}{{\tt arXiv:0806.0949}}.
\bibitem[{{Raichur} and {Paul}(2010a)}]{2010MNRAS.406.2663R}
\bibinfo{author}{{Raichur}, H.}, \bibinfo{author}{{Paul}, B.},
  \bibinfo{year}{2010}a.
\newblock \bibinfo{title}{{Apsidal motion in 4U0115+63 and orbital parameters
  of 2S1417-624 and V0332+53}}.
\newblock \bibinfo{journal}{\mnras} \bibinfo{volume}{406},
  \bibinfo{pages}{2663--2670}.
\newblock \DOIprefix\doi{10.1111/j.1365-2966.2010.16862.x}.
\bibitem[{{Raichur} and {Paul}(2010b)}]{2010MNRAS.401.1532R}
\bibinfo{author}{{Raichur}, H.}, \bibinfo{author}{{Paul}, B.},
  \bibinfo{year}{2010}b.
\newblock \bibinfo{title}{{Effect of pulse profile variations on measurement of
  eccentricity in orbits of Cen X-3 and SMC X-1}}.
\newblock \bibinfo{journal}{\mnras} \bibinfo{volume}{401},
  \bibinfo{pages}{1532--1539}.
\newblock \DOIprefix\doi{10.1111/j.1365-2966.2009.15778.x},
  \href{http://arxiv.org/abs/0909.4271}{{\tt arXiv:0909.4271}}.
\bibitem[{{Santangelo} et~al.(1998){Santangelo}, {del Sordo}, {Segreto}, {dal
  Fiume}, {Orlandini} and {Piraino}}]{1998A&A...340L..55S}
\bibinfo{author}{{Santangelo}, A.}, \bibinfo{author}{{del Sordo}, S.},
  \bibinfo{author}{{Segreto}, A.}, \bibinfo{author}{{dal Fiume}, D.},
  \bibinfo{author}{{Orlandini}, M.}, \bibinfo{author}{{Piraino}, S.},
  \bibinfo{year}{1998}.
\newblock \bibinfo{title}{{BeppoSAX detection of a Cyclotron Feature in the
  spectrum of Cen X-3}}.
\newblock \bibinfo{journal}{\aap} \bibinfo{volume}{340},
  \bibinfo{pages}{L55--L59}.
\bibitem[{{Schreier} et~al.(1972){Schreier}, {Levinson}, {Gursky}, {Kellogg},
  {Tananbaum} and {Giacconi}}]{1972ApJ...172L..79S}
\bibinfo{author}{{Schreier}, E.}, \bibinfo{author}{{Levinson}, R.},
  \bibinfo{author}{{Gursky}, H.}, \bibinfo{author}{{Kellogg}, E.},
  \bibinfo{author}{{Tananbaum}, H.}, \bibinfo{author}{{Giacconi}, R.},
  \bibinfo{year}{1972}.
\newblock \bibinfo{title}{{Evidence for the Binary Nature of Centaurus X-3 from
  UHURU X-Ray Observations.}}
\newblock \bibinfo{journal}{\apjl} \bibinfo{volume}{172}, \bibinfo{pages}{L79}.
\newblock \DOIprefix\doi{10.1086/180896}.
\bibitem[{{Shirke} et~al.(2021){Shirke}, {Bala}, {Roy} and
  {Bhattacharya}}]{2021JApA...42...58S}
\bibinfo{author}{{Shirke}, P.}, \bibinfo{author}{{Bala}, S.},
  \bibinfo{author}{{Roy}, J.}, \bibinfo{author}{{Bhattacharya}, D.},
  \bibinfo{year}{2021}.
\newblock \bibinfo{title}{{A new measurement of the spin and orbital parameters
  of the high mass X-ray binary Centaurus X-3 using AstroSat}}.
\newblock \bibinfo{journal}{Journal of Astrophysics and Astronomy}
  \bibinfo{volume}{42}, \bibinfo{pages}{58}.
\newblock \DOIprefix\doi{10.1007/s12036-021-09710-w},
  \href{http://arxiv.org/abs/2103.07204}{{\tt arXiv:2103.07204}}.
\bibitem[{{Suchy} et~al.(2008){Suchy}, {Pottschmidt}, {Wilms}, {Kreykenbohm},
  {Sch{\"o}nherr}, {Kretschmar}, {McBride}, {Caballero}, {Rothschild} and
  {Grinberg}}]{2008ApJ...675.1487S}
\bibinfo{author}{{Suchy}, S.}, \bibinfo{author}{{Pottschmidt}, K.},
  \bibinfo{author}{{Wilms}, J.}, \bibinfo{author}{{Kreykenbohm}, I.},
  \bibinfo{author}{{Sch{\"o}nherr}, G.}, \bibinfo{author}{{Kretschmar}, P.},
  \bibinfo{author}{{McBride}, V.}, \bibinfo{author}{{Caballero}, I.},
  \bibinfo{author}{{Rothschild}, R.E.}, \bibinfo{author}{{Grinberg}, V.},
  \bibinfo{year}{2008}.
\newblock \bibinfo{title}{{Pulse Phase-Resolved Analysis of the High-Mass X-Ray
  Binary Centaurus X-3 over Two Binary Orbits}}.
\newblock \bibinfo{journal}{\apj} \bibinfo{volume}{675},
  \bibinfo{pages}{1487--1498}.
\newblock \DOIprefix\doi{10.1086/527042},
  \href{http://arxiv.org/abs/0711.2752}{{\tt arXiv:0711.2752}}.
\bibitem[{{Takeshima} et~al.(1991){Takeshima}, {Dotani}, {Mitsuda} and
  {Nagase}}]{1991PASJ...43L..43T}
\bibinfo{author}{{Takeshima}, T.}, \bibinfo{author}{{Dotani}, T.},
  \bibinfo{author}{{Mitsuda}, K.}, \bibinfo{author}{{Nagase}, F.},
  \bibinfo{year}{1991}.
\newblock \bibinfo{title}{{Quasi-Periodic Oscillations in the X-Ray Flux from
  Centaurus X-3 Observed with GINGA}}.
\newblock \bibinfo{journal}{\pasj} \bibinfo{volume}{43},
  \bibinfo{pages}{L43--L50}.
\bibitem[{Tamba et~al.(2023)Tamba, Odaka, Tanimoto, Suzuki, Takashima and
  Bamba}]{Tamba_2023}
\bibinfo{author}{Tamba, T.}, \bibinfo{author}{Odaka, H.},
  \bibinfo{author}{Tanimoto, A.}, \bibinfo{author}{Suzuki, H.},
  \bibinfo{author}{Takashima, S.}, \bibinfo{author}{Bamba, A.},
  \bibinfo{year}{2023}.
\newblock \bibinfo{title}{Orbital- and spin-phase variability in the x-ray
  emission from the accreting pulsar centaurus x-3}.
\newblock \bibinfo{journal}{The Astrophysical Journal} \bibinfo{volume}{944},
  \bibinfo{pages}{9}.
\newblock \URLprefix \url{https://dx.doi.org/10.3847/1538-4357/acadde},
  \DOIprefix\doi{10.3847/1538-4357/acadde}.
\bibitem[{{Tuohy}(1976)}]{1976MNRAS.174P..45T}
\bibinfo{author}{{Tuohy}, I.R.}, \bibinfo{year}{1976}.
\newblock \bibinfo{title}{{Pulsed X-ray observations of Cen X-3 from Ariel-5.}}
\newblock \bibinfo{journal}{\mnras} \bibinfo{volume}{174},
  \bibinfo{pages}{45P--50}.
\newblock \DOIprefix\doi{10.1093/mnras/174.1.45P}.
\bibitem[{{van den Eijnden} et~al.(2021){van den Eijnden}, {Degenaar},
  {Russell}, {Wijnands}, {Bahramian}, {Miller-Jones}, {Hern{\'a}ndez
  Santisteban}, {Gallo}, {Atri}, {Plotkin}, {Maccarone}, {Sivakoff}, {Miller},
  {Reynolds}, {Russell}, {Maitra}, {Heinke}, {Armas Padilla} and
  {Shaw}}]{2021MNRAS.507.3899V}
\bibinfo{author}{{van den Eijnden}, J.}, \bibinfo{author}{{Degenaar}, N.},
  \bibinfo{author}{{Russell}, T.D.}, \bibinfo{author}{{Wijnands}, R.},
  \bibinfo{author}{{Bahramian}, A.}, \bibinfo{author}{{Miller-Jones}, J.C.A.},
  \bibinfo{author}{{Hern{\'a}ndez Santisteban}, J.V.},
  \bibinfo{author}{{Gallo}, E.}, \bibinfo{author}{{Atri}, P.},
  \bibinfo{author}{{Plotkin}, R.M.}, \bibinfo{author}{{Maccarone}, T.J.},
  \bibinfo{author}{{Sivakoff}, G.}, \bibinfo{author}{{Miller}, J.M.},
  \bibinfo{author}{{Reynolds}, M.}, \bibinfo{author}{{Russell}, D.M.},
  \bibinfo{author}{{Maitra}, D.}, \bibinfo{author}{{Heinke}, C.O.},
  \bibinfo{author}{{Armas Padilla}, M.}, \bibinfo{author}{{Shaw}, A.W.},
  \bibinfo{year}{2021}.
\newblock \bibinfo{title}{{A new radio census of neutron star X-ray binaries}}.
\newblock \bibinfo{journal}{\mnras} \bibinfo{volume}{507},
  \bibinfo{pages}{3899--3922}.
\newblock \DOIprefix\doi{10.1093/mnras/stab1995},
  \href{http://arxiv.org/abs/2107.05286}{{\tt arXiv:2107.05286}}.
\bibitem[{{van der Klis} et~al.(1980){van der Klis}, {Bonnet-Bidaud} and
  {Robba}}]{1980A}
\bibinfo{author}{{van der Klis}, M.}, \bibinfo{author}{{Bonnet-Bidaud}, J.M.},
  \bibinfo{author}{{Robba}, N.R.}, \bibinfo{year}{1980}.
\newblock \bibinfo{title}{{Characteristics of the Cen X-3 neutron star from
  correlated spin-up and X-ray tics luminosity measurements .}}
\newblock \bibinfo{journal}{\aap} \bibinfo{volume}{88}, \bibinfo{pages}{8--14}.
\bibitem[{{Wang}(2014)}]{2014MNRAS.440.1114W}
\bibinfo{author}{{Wang}, W.}, \bibinfo{year}{2014}.
\newblock \bibinfo{title}{{Variations and correlations in cyclotron resonant
  scattering features of Vela X-1 studied by INTEGRAL}}.
\newblock \bibinfo{journal}{\mnras} \bibinfo{volume}{440},
  \bibinfo{pages}{1114--1124}.
\newblock \DOIprefix\doi{10.1093/mnras/stu210},
  \href{http://arxiv.org/abs/1401.7421}{{\tt arXiv:1401.7421}}.
\bibitem[{{Wang} et~al.(2021){Wang}, {Tang}, {Tuo}, {Epili}, {Zhang}, {Song},
  {Lu}, {Qu}, {Zhang}, {Ge}, {Huang}, {Li}, {Bu}, {Cai}, {Cao}, {Chang},
  {Chen}, {Chen}, {Chen}, {Chen}, {Chen}, {Cui}, {Du}, {Gao}, {Gao}, {Gu},
  {Guan}, {Guo}, {Han}, {Huo}, {Jia}, {Jiang}, {Jin}, {Kong}, {Li}, {Li}, {Li},
  {Li}, {Li}, {Li}, {Li}, {Li}, {Liang}, {Liao}, {Liu}, {Liu}, {Liu}, {Liu},
  {Lu}, {Luo}, {Luo}, {Ma}, {Ma}, {Meng}, {Nang}, {Nie}, {Ou}, {Ren}, {Sai},
  {Song}, {Sun}, {Tao}, {Wang}, {Wang}, {Wang}, {Wang}, {Wang}, {Wen}, {Wu},
  {Wu}, {Wu}, {Xiao}, {Xiao}, {Xiong}, {Xu}, {Yang}, {Yang}, {Yang}, {Yang},
  {Yi}, {Yin}, {You}, {Zhang}, {Zhang}, {Zhang}, {Zhang}, {Zhang}, {Zhang},
  {Zhang}, {Zhang}, {Zhao}, {Zhao}, {Zheng}, {Zheng} and
  {Zhou}}]{2021JHEAp..30....1W}
\bibinfo{author}{{Wang}, W.}, \bibinfo{author}{{Tang}, Y.M.},
  \bibinfo{author}{{Tuo}, Y.L.}, \bibinfo{author}{{Epili}, P.R.},
  \bibinfo{author}{{Zhang}, S.N.}, \bibinfo{author}{{Song}, L.M.},
  \bibinfo{author}{{Lu}, F.J.}, \bibinfo{author}{{Qu}, J.L.},
  \bibinfo{author}{{Zhang}, S.}, \bibinfo{author}{{Ge}, M.Y.},
  \bibinfo{author}{{Huang}, Y.}, \bibinfo{author}{{Li}, B.},
  \bibinfo{author}{{Bu}, Q.C.}, \bibinfo{author}{{Cai}, C.},
  \bibinfo{author}{{Cao}, X.L.}, \bibinfo{author}{{Chang}, Z.},
  \bibinfo{author}{{Chen}, L.}, \bibinfo{author}{{Chen}, T.X.},
  \bibinfo{author}{{Chen}, Y.B.}, \bibinfo{author}{{Chen}, Y.},
  \bibinfo{author}{{Chen}, Y.P.}, \bibinfo{author}{{Cui}, W.W.},
  \bibinfo{author}{{Du}, Y.Y.}, \bibinfo{author}{{Gao}, G.H.},
  \bibinfo{author}{{Gao}, H.}, \bibinfo{author}{{Gu}, Y.D.},
  \bibinfo{author}{{Guan}, J.}, \bibinfo{author}{{Guo}, C.C.},
  \bibinfo{author}{{Han}, D.W.}, \bibinfo{author}{{Huo}, J.},
  \bibinfo{author}{{Jia}, S.M.}, \bibinfo{author}{{Jiang}, W.C.},
  \bibinfo{author}{{Jin}, J.}, \bibinfo{author}{{Kong}, L.D.},
  \bibinfo{author}{{Li}, C.K.}, \bibinfo{author}{{Li}, G.},
  \bibinfo{author}{{Li}, T.P.}, \bibinfo{author}{{Li}, W.},
  \bibinfo{author}{{Li}, X.}, \bibinfo{author}{{Li}, X.B.},
  \bibinfo{author}{{Li}, X.F.}, \bibinfo{author}{{Li}, Z.W.},
  \bibinfo{author}{{Liang}, X.H.}, \bibinfo{author}{{Liao}, J.Y.},
  \bibinfo{author}{{Liu}, B.S.}, \bibinfo{author}{{Liu}, C.Z.},
  \bibinfo{author}{{Liu}, H.X.}, \bibinfo{author}{{Liu}, H.W.},
  \bibinfo{author}{{Lu}, X.F.}, \bibinfo{author}{{Luo}, Q.},
  \bibinfo{author}{{Luo}, T.}, \bibinfo{author}{{Ma}, R.C.},
  \bibinfo{author}{{Ma}, X.}, \bibinfo{author}{{Meng}, B.},
  \bibinfo{author}{{Nang}, Y.}, \bibinfo{author}{{Nie}, J.Y.},
  \bibinfo{author}{{Ou}, G.}, \bibinfo{author}{{Ren}, X.Q.},
  \bibinfo{author}{{Sai}, N.}, \bibinfo{author}{{Song}, X.Y.},
  \bibinfo{author}{{Sun}, L.}, \bibinfo{author}{{Tao}, L.},
  \bibinfo{author}{{Wang}, C.}, \bibinfo{author}{{Wang}, L.J.},
  \bibinfo{author}{{Wang}, P.J.}, \bibinfo{author}{{Wang}, W.S.},
  \bibinfo{author}{{Wang}, Y.S.}, \bibinfo{author}{{Wen}, X.Y.},
  \bibinfo{author}{{Wu}, B.Y.}, \bibinfo{author}{{Wu}, B.B.},
  \bibinfo{author}{{Wu}, M.}, \bibinfo{author}{{Xiao}, G.C.},
  \bibinfo{author}{{Xiao}, S.}, \bibinfo{author}{{Xiong}, S.L.},
  \bibinfo{author}{{Xu}, Y.P.}, \bibinfo{author}{{Yang}, R.J.},
  \bibinfo{author}{{Yang}, S.}, \bibinfo{author}{{Yang}, J.J.},
  \bibinfo{author}{{Yang}, Y.J.}, \bibinfo{author}{{Yi}, B.B.},
  \bibinfo{author}{{Yin}, Q.Q.}, \bibinfo{author}{{You}, Y.},
  \bibinfo{author}{{Zhang}, F.}, \bibinfo{author}{{Zhang}, H.M.},
  \bibinfo{author}{{Zhang}, J.}, \bibinfo{author}{{Zhang}, P.},
  \bibinfo{author}{{Zhang}, W.}, \bibinfo{author}{{Zhang}, W.C.},
  \bibinfo{author}{{Zhang}, Y.F.}, \bibinfo{author}{{Zhang}, Y.H.},
  \bibinfo{author}{{Zhao}, H.S.}, \bibinfo{author}{{Zhao}, X.F.},
  \bibinfo{author}{{Zheng}, S.J.}, \bibinfo{author}{{Zheng}, Y.G.},
  \bibinfo{author}{{Zhou}, D.K.}, \bibinfo{year}{2021}.
\newblock \bibinfo{title}{{Accretion torque reversals in GRO J1008-57 revealed
  by Insight-HXMT}}.
\newblock \bibinfo{journal}{Journal of High Energy Astrophysics}
  \bibinfo{volume}{30}, \bibinfo{pages}{1--8}.
\newblock \DOIprefix\doi{10.1016/j.jheap.2021.01.002},
  \href{http://arxiv.org/abs/2102.12085}{{\tt arXiv:2102.12085}}.
\bibitem[{{Zhang} et~al.(2020){Zhang}, {Li}, {Lu}, {Song}, {Xu}, {Liu}, {Chen},
  {Cao}, {Bu}, {Chang}, {Chen}, {Chen}, {Chen}, {Chen}, {Chen}, {Cui}, {Cui},
  {Deng}, {Dong}, {Du}, {Fu}, {Gao}, {Gao}, {Gao}, {Ge}, {Gu}, {Guan},
  {Gungor}, {Guo}, {Han}, {Hu}, {Huang}, {Huo}, {Jia}, {Jiang}, {Jiang}, {Jin},
  {Jin}, {Li}, {Li}, {Li}, {Li}, {Li}, {Li}, {Li}, {Li}, {Li}, {Li}, {Li},
  {Liang}, {Liao}, {Liu}, {Liu}, {Liu}, {Liu}, {Liu}, {Liu}, {Lu}, {Lu}, {Luo},
  {Ma}, {Meng}, {Nang}, {Nie}, {Ou}, {Qu}, {Sai}, {Shang}, {Shen}, {Sun},
  {Tan}, {Tao}, {Tuo}, {Wang}, {Wang}, {Wang}, {Wang}, {Wang}, {Wang}, {Wang},
  {Wen}, {Wu}, {Wu}, {Wu}, {Xiao}, {Xiong}, {Yan}, {Yang}, {Yang}, {Yang},
  {Yi}, {Yuan}, {Zhang}, {Zhang}, {Zhang}, {Zhang}, {Zhang}, {Zhang}, {Zhang},
  {Zhang}, {Zhang}, {Zhang}, {Zhang}, {Zhang}, {Zhang}, {Zhang}, {Zhang},
  {Zhang}, {Zhang}, {Zhang}, {Zhang}, {Zhang}, {Zhao}, {Zhao}, {Zheng}, {Zhou},
  {Zhu}, {Zhu}, {Zhuang} and {Insight-HXMT Team}}]{2020SCPMA..6349502Z}
\bibinfo{author}{{Zhang}, S.N.}, \bibinfo{author}{{Li}, T.},
  \bibinfo{author}{{Lu}, F.}, \bibinfo{author}{{Song}, L.},
  \bibinfo{author}{{Xu}, Y.}, \bibinfo{author}{{Liu}, C.},
  \bibinfo{author}{{Chen}, Y.}, \bibinfo{author}{{Cao}, X.},
  \bibinfo{author}{{Bu}, Q.}, \bibinfo{author}{{Chang}, Z.},
  \bibinfo{author}{{Chen}, G.}, \bibinfo{author}{{Chen}, L.},
  \bibinfo{author}{{Chen}, T.}, \bibinfo{author}{{Chen}, Y.},
  \bibinfo{author}{{Chen}, Y.}, \bibinfo{author}{{Cui}, W.},
  \bibinfo{author}{{Cui}, W.}, \bibinfo{author}{{Deng}, J.},
  \bibinfo{author}{{Dong}, Y.}, \bibinfo{author}{{Du}, Y.},
  \bibinfo{author}{{Fu}, M.}, \bibinfo{author}{{Gao}, G.},
  \bibinfo{author}{{Gao}, H.}, \bibinfo{author}{{Gao}, M.},
  \bibinfo{author}{{Ge}, M.}, \bibinfo{author}{{Gu}, Y.},
  \bibinfo{author}{{Guan}, J.}, \bibinfo{author}{{Gungor}, C.},
  \bibinfo{author}{{Guo}, C.}, \bibinfo{author}{{Han}, D.},
  \bibinfo{author}{{Hu}, W.}, \bibinfo{author}{{Huang}, Y.},
  \bibinfo{author}{{Huo}, J.}, \bibinfo{author}{{Jia}, S.},
  \bibinfo{author}{{Jiang}, L.}, \bibinfo{author}{{Jiang}, W.},
  \bibinfo{author}{{Jin}, J.}, \bibinfo{author}{{Jin}, Y.},
  \bibinfo{author}{{Li}, B.}, \bibinfo{author}{{Li}, C.},
  \bibinfo{author}{{Li}, G.}, \bibinfo{author}{{Li}, M.},
  \bibinfo{author}{{Li}, W.}, \bibinfo{author}{{Li}, X.},
  \bibinfo{author}{{Li}, X.}, \bibinfo{author}{{Li}, X.},
  \bibinfo{author}{{Li}, Y.}, \bibinfo{author}{{Li}, Z.},
  \bibinfo{author}{{Li}, Z.}, \bibinfo{author}{{Liang}, X.},
  \bibinfo{author}{{Liao}, J.}, \bibinfo{author}{{Liu}, G.},
  \bibinfo{author}{{Liu}, H.}, \bibinfo{author}{{Liu}, S.},
  \bibinfo{author}{{Liu}, X.}, \bibinfo{author}{{Liu}, Y.},
  \bibinfo{author}{{Liu}, Y.}, \bibinfo{author}{{Lu}, B.},
  \bibinfo{author}{{Lu}, X.}, \bibinfo{author}{{Luo}, T.},
  \bibinfo{author}{{Ma}, X.}, \bibinfo{author}{{Meng}, B.},
  \bibinfo{author}{{Nang}, Y.}, \bibinfo{author}{{Nie}, J.},
  \bibinfo{author}{{Ou}, G.}, \bibinfo{author}{{Qu}, J.},
  \bibinfo{author}{{Sai}, N.}, \bibinfo{author}{{Shang}, R.},
  \bibinfo{author}{{Shen}, G.}, \bibinfo{author}{{Sun}, L.},
  \bibinfo{author}{{Tan}, Y.}, \bibinfo{author}{{Tao}, L.},
  \bibinfo{author}{{Tuo}, Y.}, \bibinfo{author}{{Wang}, C.},
  \bibinfo{author}{{Wang}, C.}, \bibinfo{author}{{Wang}, G.},
  \bibinfo{author}{{Wang}, H.}, \bibinfo{author}{{Wang}, J.},
  \bibinfo{author}{{Wang}, W.}, \bibinfo{author}{{Wang}, Y.},
  \bibinfo{author}{{Wen}, X.}, \bibinfo{author}{{Wu}, B.},
  \bibinfo{author}{{Wu}, B.}, \bibinfo{author}{{Wu}, M.},
  \bibinfo{author}{{Xiao}, G.}, \bibinfo{author}{{Xiong}, S.},
  \bibinfo{author}{{Yan}, L.}, \bibinfo{author}{{Yang}, J.},
  \bibinfo{author}{{Yang}, S.}, \bibinfo{author}{{Yang}, Y.},
  \bibinfo{author}{{Yi}, Q.}, \bibinfo{author}{{Yuan}, B.},
  \bibinfo{author}{{Zhang}, A.}, \bibinfo{author}{{Zhang}, C.},
  \bibinfo{author}{{Zhang}, C.}, \bibinfo{author}{{Zhang}, F.},
  \bibinfo{author}{{Zhang}, H.}, \bibinfo{author}{{Zhang}, J.},
  \bibinfo{author}{{Zhang}, Q.}, \bibinfo{author}{{Zhang}, S.},
  \bibinfo{author}{{Zhang}, S.}, \bibinfo{author}{{Zhang}, T.},
  \bibinfo{author}{{Zhang}, W.}, \bibinfo{author}{{Zhang}, W.},
  \bibinfo{author}{{Zhang}, W.}, \bibinfo{author}{{Zhang}, Y.},
  \bibinfo{author}{{Zhang}, Y.}, \bibinfo{author}{{Zhang}, Y.},
  \bibinfo{author}{{Zhang}, Y.}, \bibinfo{author}{{Zhang}, Z.},
  \bibinfo{author}{{Zhang}, Z.}, \bibinfo{author}{{Zhang}, Z.},
  \bibinfo{author}{{Zhao}, H.}, \bibinfo{author}{{Zhao}, X.},
  \bibinfo{author}{{Zheng}, S.}, \bibinfo{author}{{Zhou}, J.},
  \bibinfo{author}{{Zhu}, Y.}, \bibinfo{author}{{Zhu}, Y.},
  \bibinfo{author}{{Zhuang}, R.}, \bibinfo{author}{{Insight-HXMT Team}},
  \bibinfo{year}{2020}.
\newblock \bibinfo{title}{{Overview to the Hard X-ray Modulation Telescope
  (Insight-HXMT) Satellite}}.
\newblock \bibinfo{journal}{Science China Physics, Mechanics, and Astronomy}
  \bibinfo{volume}{63}, \bibinfo{pages}{249502}.
\newblock \DOIprefix\doi{10.1007/s11433-019-1432-6},
  \href{http://arxiv.org/abs/1910.09613}{{\tt arXiv:1910.09613}}.
\bibitem[{{Zhang} et~al.(2001){Zhang}, {Marshall}, {Gotthelf}, {Middleditch}
  and {Wang}}]{2001ApJ...554L.177Z}
\bibinfo{author}{{Zhang}, W.}, \bibinfo{author}{{Marshall}, F.E.},
  \bibinfo{author}{{Gotthelf}, E.V.}, \bibinfo{author}{{Middleditch}, J.},
  \bibinfo{author}{{Wang}, Q.D.}, \bibinfo{year}{2001}.
\newblock \bibinfo{title}{{A Phase-connected Braking Index Measurement for the
  Large Magellanic Cloud Pulsar PSR B0540-69}}.
\newblock \bibinfo{journal}{\apjl} \bibinfo{volume}{554},
  \bibinfo{pages}{L177--L180}.
\newblock \DOIprefix\doi{10.1086/321703}.

\end{thebibliography}

\end{document}